\documentclass[journal]{IEEEtran}
\IEEEoverridecommandlockouts
\usepackage{cite}
\usepackage{amsmath,amssymb,amsfonts}
\usepackage{algorithmic}
\usepackage{graphicx}
\usepackage{textcomp}
\usepackage{subfigure}
\usepackage{float}
\usepackage{xcolor}
\usepackage{enumerate}
\usepackage[justification=centering]{caption}

\usepackage[normalem]{ulem}
\usepackage{cancel}
\usepackage{tikz}

\usepackage{cite}
\usepackage{amsmath,amssymb,amsfonts}
\usepackage{microtype}
\usepackage{algorithmic}
\usepackage{graphicx}
\usepackage{textcomp}
\usepackage{subfigure}
\usepackage{float}
\usepackage{xcolor}
\usepackage{array}
\usepackage{enumerate}
\newtheorem{theorem}{Theorem}
\newtheorem{proposition}{Proposition}
\newtheorem{lemma}{Lemma}
\newtheorem{definition}{Definition}

\newtheorem{remark}{Remark}

\usepackage{geometry}

\newcommand{\Pe}{P_e}

\allowdisplaybreaks[4]
\newcommand{\sgn}[1]{\mathrm{sgn}(#1)}
\begin{document}
	\title{Coding for Fading Channels with Imperfect CSI at the Transmitter and Quantized Feedback}
	\author{ Yuhan~Yang, Haoheng~Yuan, Chao~Qi, Fan~Cheng and Bin~Dai
	}
	\maketitle
	\begin{abstract}
		In wireless communication systems, the intended receiver is able to obtain the perfect channel state information (CSI) as long as the training sequence is sufficiently long, and then through a quantized feedback channel (QFC), the transmitter gets imperfect CSI caused by the quantized noise. In general, the design of efficient coding schemes for wireless channels with imperfect CSI at the transmitter (I-CSIT) is difficult and challenging, and one possible method is to construct such a scheme by using channel feedback. In the literature, it has already been shown that the classical Schalkwijk-Kailath (SK) scheme for the additive Gaussian noise channel with noiseless feedback is a highly efficient coding scheme since its coding complexity is extremely low and the decoding error doubly exponentially decays as the coding blocklength tends to infinity. However, the SK-type schemes are very sensitive to the CSI and channel output feedback, which indicates that extending these schemes to channels with I-CSIT and QFC is challenging. In this paper, first, for the quasi-static fading channel with I-CSIT and QFC, we show that introducing modulo lattice function and an auxiliary signal into the SK-type encoder-decoder of the transceiver, the decoding error caused by the imperfect CSI at the transmitter can be perfectly eliminated, resulting in the success of designing SK-type scheme for such a case. Next, we further extend the above scheme to 
		the two-path fading scenario, which is modeled as the two-ray channel with I-CSIT and QFC.
		Treating the signal of the second path as a relay, our extended scheme combines the amplify-and-forward relay strategy and the previously proposed scheme for the same model without ISI. {However, this extended scheme cannot be applied to multi-path fading scenario, to this end,
			we propose a new SK-type scheme for the multi-path fading channel with noiseless feedback and perfect CSI at the transceiver. The intuition behind this scheme is to transform 
			the multi-path fading channel into a fading MIMO channel by discrete fourier transform (DFT), and then apply the SK-type scheme to the MIMO channel in frequency domain and use inverse DFT
			to obtain the codeword in time domain. The study of this paper may provide a way to design efficient coding scheme for multi-path fading channels in the presence of imperfect CSI and quantized feedback.}
		
	\end{abstract}
	\section{Introduction}\label{sec1}
	
	In modern wireless systems,
	ultra-reliable low-latency communication (URLLC) \cite{fbl2,fbl3,urllc1,urllc2}
	plays a key role since it supports many
	critical services requiring high reliability and
	low latency, such as unmanned aerial vehicle (UAV) communication network \cite{UAV,UAV1,UAV2}, Vehicle-to-Everything (V2X) communication \cite{V2X,V2X1,V2X2}, etc, and finite blocklength (FBL) coding \cite{fbl1} provides a way to achieve it.
	
	On the other hand, in practical wireless communication channels,
	it is possible for the receiver to obtain perfect channel state information (CSI) provided that the average power of pilot symbols is sufficiently high or the pilot length is sufficiently long \cite{H_est}, and the transmitter is access to the receiver's obtained CSI via a quantized feedback channel (QFC), which indicates that
	the transmitter can only obtain imperfect CSI.
	In such a case,
	though existing FBL codes (e.g., LDPC and Polar codes) can still work by choosing
	much lower transmission rate \cite{3GPP}, these schemes are far from optimal since the imperfect CSI at the transmitter (I-CSIT) is not fully exploited.

	Half a century ago, a highly effective coding scheme, known as the Schalkwijk-Kailath (SK) scheme \cite{JS7} was proposed,
	which is based on the perfect channel feedback and additive white Gaussian noise (AWGN) channel. In such a scheme, at the beginning, the message is un-coded delivered to the receiver, and
	the receiver does his first estimation about the message by normalizing his received signal.
	Through perfect channel feedback, the transmitter also knows the receiver's estimation, and computes the estimation error by subtracting the real message from the estimation. In next time instants,
	the transmitter forwards the receiver's estimation error at previous time, and the receiver adopts minimum mean square estimation (MMSE) about this estimation error
	once receiving the fresh signal, which is used to update the receiver's message estimation.
	By iteration, it is shown that the receiver's message estimation error vanishes as the coding blocklength tends to infinity, and the transmission rate of this scheme approaches the capacity of the AWGN channel when the blocklength tends to infinity\footnote{In fact, the SK scheme also almost approaches the FBL capacity of the AWGN channel with feedback \cite{fbl1}.}.
	Besides this, comparing with the well-known linear block codes, the decoding error probability (DEP) of the SK scheme decreases as a second-order exponential in the coding blocklength, which indicates that to achieve a desired DEP, the coding blocklength of the SK scheme is much shorter.
	
	
{Though the SK scheme performs well, its application to practical communication scenarios is still a big challenge due to the noiseless feedback assumption. In \cite{JS10}, it has been shown that when the feedback channel is also an AWGN channel, there exists significant performance loss in the classical SK scheme \cite{JS7}. Introducing modulo-lattice function (MLF) into the iteration stage of the SK scheme, \cite{JS7} showed that the MLF based SK scheme almost approaches the transmission rate of the classical SK scheme when the power of the feedback encoder is sufficiently large. However, note that the SK-type scheme of \cite{JS7} cannot be extended to the quantized feedback case, where the feedback channel is noise-free and the feedback signal is quantized by the receiver
		which indicates that the receiver knows the quantized noise in a causal manner, and the reason is that the receiver's MMSE co-efficient factor cannot be computed since the probability distribution of the quantized noise is unknown to all. To deal with this, \cite{JS11} modified the SK scheme by determining the coding parameters via a linear quadratic Gaussian (LQG) optimization tool \cite{LQG}.
		Another challenge to the application of SK-type schemes is feedback delay. In \cite{Markov}-\cite{mar3}, optimal SK-type schemes for
		the finite state Markov channel with delayed feedback were proposed, where the transmitted message is split into sub-messages according to the feedback delay time, and each sub-message
		is encoded by an SK-type scheme with water-filling transmitting power.
		Very recently, the SK scheme has also been extended to multiple-input multiple-output (MIMO) systems \cite{MIMO}-\cite{xie}, where the MIMO channel is divided into several independent sub-channels, and the overall message is split into the same number of sub-messages with each sub-message encoded by an SK-type scheme and transmitted over a sub-channel.}
	
	{As stated above, though the SK-type schemes are well developed in various feedback setting and channel models, however, note that all above mentioned works are based on the assumption that both the transmitter and the receiver perfectly know the channel state information (CSI), which is a big challenge in practical wireless communication scenarios. On the other hand, if the transmitter only knows the imperfect CSI, the estimation error of CSI will involved in the iteration of SK-type schemes, which leads to the receiver's final estimation about the transmitted message does not converge when the coding blocklength tends to infinity, and hence how to design SK-type scheme under such imperfect CSI case is of great importance.}

	In this paper, we aim to extend the SK-type scheme in the literature to the wireless fading channel with I-CSIT and QFC. First, for the quasi-static fading channel, we introduce modulo lattice function and an auxiliary signal into the SK-type encoder-decoder of the transceiver, and show that the decoding error caused by the imperfect CSI at the transmitter can be perfectly eliminated, resulting in the success of designing SK-type scheme for such a case. Next, we further extend the above scheme to intersymbol interference (ISI) situation, namely, the two-ray channel with I-CSIT and QFC, which treats the signal of the second path as a relay, and combines the amplify-and-forward (AF) relay strategy \cite{AF} with the previously proposed scheme of the same model without ISI. {However, note that the relay based SK-type scheme cannot be further extended to multi-path fading scenario since the coding parameters determined by a series of equations may not have solutions simultaneously.
		To this end, we propose a new SK-type scheme for the multi-path fading channel with noiseless feedback and perfect CSI at the transceiver. The intuition behind this scheme is to transform
		the multi-path fading channel into a fading MIMO channel by discrete fourier transform (DFT), and then apply the SK-type scheme to the MIMO channel in frequency domain and use inverse DFT
		to obtain the codeword in time domain.} Finally, the results of this paper are further illustrated by numerical examples.
	The remainder of this paper is organized as follows. Some useful tools used in this paper are formally introduced in Section \ref{Preliminaries}. The quasi-static fading channel with I-CSIT and QFC is formulated and studied in Section \ref{sk1}. Section \ref{sk2} further considers a two-path case of the model presented in Section \ref{sk1}. {Section \ref{sk3} investigates the multi-path fading channel with noiseless feedback and perfect CSI at the transceiver.}
	Summary and future work are provided in
	Section \ref{Conclusion}.

	\emph{Notations}:
	Throughout this paper,
	the random variable (RV) is denoted by capital letter and
	its realization and range by lowercase and calligraphic letters,
	respectively.
	Random vector and its realization are denoted in the same way.
	For example, $\boldsymbol{X}^{N}$ represents a $N$-dimensional random vector $[X_{1},...,X_{N}]^{T}$, and for simplicity it is denoted by $\boldsymbol{X}$.
	$\boldsymbol{X}_i$ and $\boldsymbol{X}_{i}^{j}$ represent the $i$-th element and the \(i\)-th to the \(j\)-th elements of the vector \(\boldsymbol{X}\) $(\boldsymbol{X}_{i}^{j}=[X_i,X_{i+1},\cdots,X_{j}]^{T})$, respectively. If $i = 1$, we also use $\boldsymbol{X}^{j}$ instead of $\boldsymbol{X}_{1}^{j}$.
	$\boldsymbol{x}^{N}=(x_{1},...,x_{N})$ represents a vector value in $\mathcal{X}^{N}$ (the $N$-th Cartesian power of the alphabet $\mathcal{X}$).
	\(\boldsymbol{a}=[a_1, a_2, \dots, a_M]^{T}\) and \(\boldsymbol{b}=[b_1, b_2, \dots, b_N]^{T}\) are column vectors of dimension \(M\) and \(N\), respectively.
	Then \(\boldsymbol{C}=[\boldsymbol{a}\,;\,\boldsymbol{b}]=[a_1, a_2, \dots, a_M, b_1, b_2, \dots, b_N]^{T}\) represents a column vector of dimension \(M+N\).
	$\boldsymbol{A}_{M \times N}$ represents a
	matrix with $M$ rows and $N$ columns, and for simplicity it is denoted by $\boldsymbol{A}$.
	$\boldsymbol{A}_{i,j}$ represents the $i$-th row, $j$-th column element of matrix $\boldsymbol{A}$.  $\boldsymbol{I}_{N}$ is a $N \times N$ identity matrix.
	The $K\times K$ circulant channel matrix $\boldsymbol{A}$ with first column $[a_1, a_2, \dots, a_K]^{T}$ is defined as
	\[
	\boldsymbol{A} =
	\begin{bmatrix}
		a_1 & a_K & a_{K-1} & \cdots & a_2 \\
		a_2 & a_1 & a_K & \cdots & a_3 \\
		a_3 & a_2 & a_1 & \cdots & a_4 \\
		\vdots & \vdots & \vdots & \ddots & \vdots \\
		a_K & a_{K-1} & a_{K-2} & \cdots & a_1
	\end{bmatrix},
	\]
	$\text{diag}(x_1,\dots,x_n)$ represents the $n \times n$ diagonal matrix with entries $x_1, \dots, x_n$ on its main diagonal ($\text{diag}(x_1,\cdots,x_n)=\begin{bmatrix}
		x_1& &\boldsymbol{0} \\
		& \ddots&        \\
		\boldsymbol{0}& &x_n\\		
	\end{bmatrix}$). Let \( z \in \mathbb{C} \) be a complex number, then its real and imaginary parts are defined as \( \text{Re}(z) \) and \( \text{Im}(z) \) respectively, where \( z = \text{Re}(z) + j\text{Im}(z) \) $(j \triangleq \sqrt{-1})$, and denote its complex conjugate by \( \overline{z} \) ($\overline{z}=\text{Re}(z) - j\text{Im}(z)$). The superscript $(\cdot)^{T}$ and $(\cdot)^{H}$ represent the transpose and conjugate transpose, respectively.
	$\|\boldsymbol{x}\|$ denotes the $l_{2}$-norm of the vector $\boldsymbol{x}$ ($\|\boldsymbol{x}\|=\sqrt{\boldsymbol{x}^{H}\boldsymbol{x}}$),
	$|\mathcal{W}|$ represents the cardinality of a set $\mathcal{W}$.
	$\mathbb{E}[\cdot]$ represents statistical expectation for random variables.
	Let $\mathcal{N}(0,\sigma^{2})$ be Gaussian distribution with mean $0$ and covariance $\sigma^{2}$, and $\sim$ stands for ``distributed as''.
	The complementary Gaussian cumulative distribution function (also called $Q$-function) is defined by  $Q(x) = \frac{1}{\sqrt{2 \pi}} \int_{x}^{\infty} {\rm exp}( {\frac{-t^{2}}{2}}) dt$,
	and $Q^{-1}(\cdot)$ is the inverse function.
	$\left\lfloor x \right\rfloor\triangleq\max\left\lbrace n\in \mathbb{N}: x\geq n\right\rbrace $ and $[x]^{+}$ represents the positive part of $x$ ($[x]^{+} = \max(0, x)$). In the remainder of this paper, the base of the $\log$ function is $2$.

	\section{Preliminaries}\label{Preliminaries}
	\subsection{The SK-type scheme for the quasi-static fading channel with perfect CSI at the transceiver and noiseless feedback}\label{sk0}
	\begin{figure}[htb]
		\centering
		\includegraphics[scale=0.16]{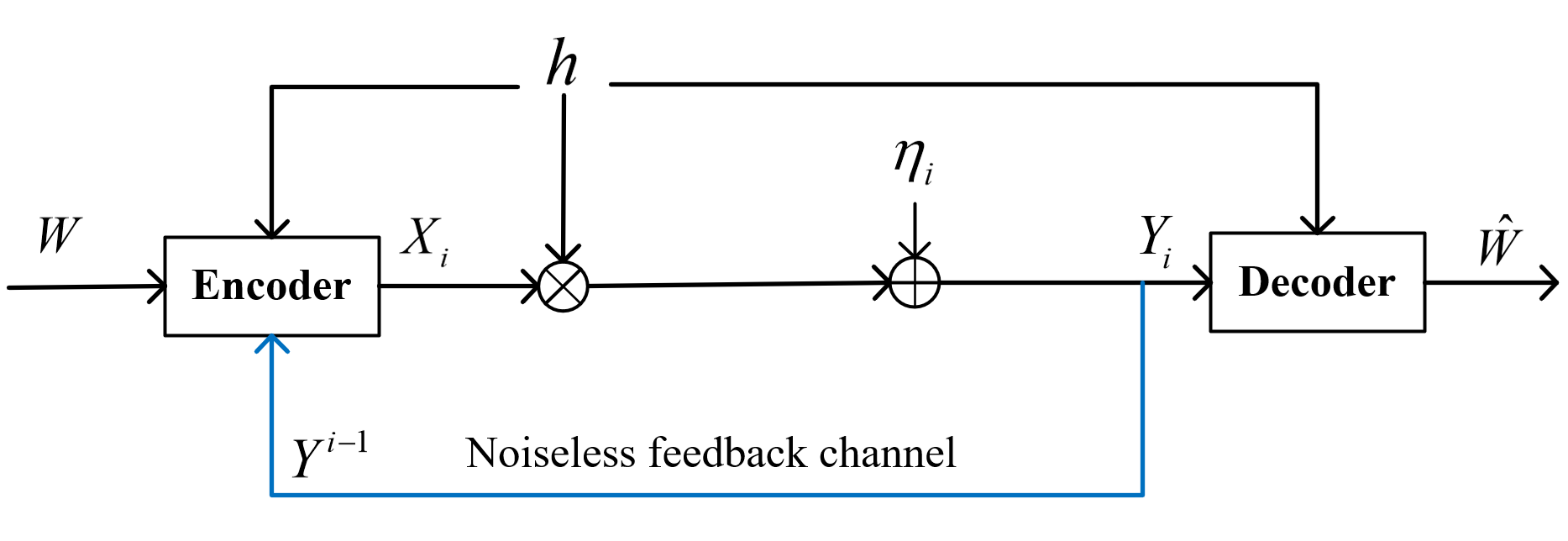}
		\caption{The quasi-static fading channel with perfect CSI at the transceiver (P-CSI-TR) and noiseless feedback (NF)}
		\label{fig0}
	\end{figure}
	In Figure \ref{fig0}, a transmitter wishes to send a message $W$, which is uniformly drawn from $\mathcal{W} = \{1,2,\ldots,2^{NR}\}$, to a receiver, and
	at time $i\in\left\lbrace 1,2,\dots,N\right\rbrace $, the channel  output is characterized by
	{\small 	\begin{equation}\label{model}
			Y_i=hX_i+\eta_i,
			\vspace{-5pt}	
	\end{equation}	}%
	where $h$ is the quasi-static fading coefficient, namely, $h$  remains a constant within a block, and it is perfectly known by the transceiver,
	$X_i=g_i(W,\boldsymbol{Y}^{i-1},h)$ is the channel input at time $i$ and it satisfies an average power constraint $\frac{1}{N}\sum\nolimits_{i=1}^{N}\mathbb{E}[{X}_{i}^2]\leq P$,
	and $\eta_i\sim\mathcal{N}(0,\sigma^{2})$ is an AWGN which is independent and identically distributed (i.i.d.) across time index $i$.
	
	The signal-to-noise ratio (SNR) is defined as $	\text{SNR}\triangleq\frac{P}{\sigma^2}$.
	
	At time $N$, the output of the decoder is $\hat{W}=\psi(\boldsymbol{Y}^{N},h)$, and the decoding error probability is defined as $P_{e}=\Pr\{\hat{W}\neq W\}$.
	
	A rate $R$ is said to be achievable if for any $\epsilon>0$ and sufficiently large $N$,
	there exists  a pair of encoder-decoder such that
	$\frac{1}{N}\log |\mathcal{W}| \geq {R}-\varepsilon$ and $P_{e}\leq \varepsilon$.
	The capacity $\mathcal{C}_\text{fd}$ of the quasi-static fading channel with P-CSI-TR and NF \cite{gamal}
	is the maximum achievable $R$, and it is given by
	{\small 	\begin{equation}\label{zzertong}
			\mathcal{C}_\text{fd}=\frac{1}{2}\log \left(1+h^2\cdot\text{SNR}\right).
			\vspace{-3pt}
	\end{equation}}%
	
	It has been already shown that an SK-type scheme achieves $\mathcal{C}_\text{fd}$ \cite{xie}, and the detail of this scheme is described below.	
	
	\subsubsection{ Message mapping}
	
	Since $\mathcal{W}=\{1,2,\cdots,2^{NR}\}$, partitioning the interval $[-0.5,0.5]$ into {\small $2^{NR}$}
	sub-intervals with equal space, and each sub-interval's
	midpoint $\theta$ is the mapping value of $W$, namely,
	{\small 	\begin{equation}\label{mapping}
			\theta=-\frac{1}{2}+\frac{2W-1}{2\cdot2^{NR}}.
	\end{equation}}%
	Since $W$ is uniformly drawn, it is not difficult to check that $\mathbb{E}[\theta^{2}]\approx\frac{1}{12}$.
	
	\emph{Initialization}: At time instant $1$,  the transmitter  encodes $\theta$ as 	$X_{1}=\sqrt{{12P}}\theta$.
	
	The receiver receives $Y_{1}=hX_{1}+\eta_{1}$, and transmits $Y_{1}$ back to the transmitter via NF.
	Since the receiver knows $h$, he computes his first estimation $\hat{\theta}_{1}$ of $\theta$ by
	{\small 	\begin{equation}
			\hat{\theta}_1=\frac{Y_1}{h\sqrt{12P}}=\theta+\frac{\eta_{1}}{h\sqrt{12P}}=\theta+\epsilon_1,
	\end{equation}}%
	where $\epsilon_1=\hat{\theta}_1-\theta$, and $\epsilon_1\sim\mathcal{N}(0,\frac{\sigma^2}{12h^{2}P})$.
	
	\subsubsection{Iteration}
	At time instant $i \in \left\lbrace {2},\ldots,N\right\rbrace$, since the transmitter knows $h$, he computes the receiver's previous estimation error $\epsilon_{i-1}=\hat{\theta}_{i-1}-\theta$, and sends
	$X_i=\sqrt{\frac{P}{\mathbb{E}[\epsilon_{i-1}^{2}]}}\epsilon_{i-1}$.
	
	Once receiving $Y_{i}=hX_{i}+\eta_{i}$,
	the receiver updates his estimation  $\hat{\theta}_{i}$ of $\theta$ by
	\vspace{-3pt}
	{\small  	\begin{equation}
			\hat{\theta}_{i}=\hat{\theta}_{i-1}-\beta_{i-1}\cdot\frac{{Y}_{i}}{h}=\theta+\epsilon_i,
			\vspace{-5pt}
	\end{equation}}%
	where  $\beta_{i-1}$ is the MMSE coefficient used to estimate $\epsilon_{i-1}$ from $\frac{Y_i}{h}$, and it is given by
	
	{\small 	\begin{equation}
			\beta_{i-1}=\frac{\mathbb{E}[\epsilon_{i-1}\cdot\frac{{Y}_{i}}{h}]}{\mathbb{E}[(\frac{{Y}_{i}}{h})^{2}]}
			=\frac{\sqrt{P\cdot\mathbb{E}[\epsilon_{i-1}^2]}}{P+\frac{\sigma^2}{h^2}}.
	\end{equation}}%
	\subsubsection{Decoding}
	After $N$ rounds of iteration, the receiver's
	final estimation of $\theta$ is $\hat{\theta}_{N}=\theta+\epsilon_{N}$. Choose the
	closest midpoint to  $\hat{\theta}_{N}$, and obtain the decoded message
	$\hat{W}$ by the mapping relationship (\ref{mapping}). It is easy to see that
	$\hat{W}$ equals to $W$ when $\hat{\theta}_{N}$ belongs to the sub-interval
	corresponding to $W$, which means the receiver  decodes
	the transmitted message correctly.

	\subsubsection{Analysis of the probability of error}\hfil \\
	The receiver's decoding error probability is given by
	{\small 	\begin{equation}\label{sk0-pe}
			P_e=\Pr\lbrace |\epsilon_{N}|\geq\frac{1}{2^{NR+1}}\rbrace
			=2\mathop{Q}( \frac{1}{2^{NR+1}\sqrt{\mathbb{E}[\epsilon_{N}^{2}]}}),
	\end{equation}}%
	where {\small $\mathop{Q}(x)\triangleq \frac{1}{\sqrt{2 \pi}} \int_{x}^{\infty} {\rm exp}( {-\frac{t^{2}}{2}}) dt$}.
	
	Here note that
	{\small 	\begin{equation}\label{sigma_p}
			\mathbb{E}[\epsilon_{i+1}^2]\!=\!\mathbb{E}[\epsilon_{i}^2]\!-\!\frac{(\mathbb{E}[\epsilon_{i}\frac{{Y}_{i+1}}{h}])^2}{\mathbb{E}[(\frac{{Y}_{i+1}}{h})^{2}]}	
			=\frac{1}{12 h^2 \text{SNR}(1+h^2\text{SNR})^{i}},
	\end{equation}}%
	Substituting $\mathbb{E}[\epsilon_{N}^2]$ into (\ref{sk0-pe}),
	{\small \begin{align}\label{sk0-pe-1}
			&P_e= 2Q( \frac{\sqrt{12\cdot h^2\cdot \text{SNR}(1+h^2\cdot\text{SNR})^{N-1}}}{2^{NR+1}})\nonumber\\
			&\quad\mathop{\leq}\limits^{\rm{(a)}}\exp ( -\frac{3}{2} \cdot\frac{ h^2\cdot\text{SNR}}{1+h^{2}\cdot\text{SNR}}\cdot2^{2N\cdot\left( \frac{1}{2}\log(1+h^2\cdot\text{SNR})-R\right) }),
	\end{align}}%
	where $\exp(x)\triangleq e^{x}$, and $Q(x)$-function is the complementary Gaussian cumulative distribution function, and (a) follows from 	$Q(x)$-function is upper bounded by $Q(x)\leq\frac{1}{2} e^{-\frac{1}{2}x^{2}}$.   (\ref{sk0-pe-1}) demonstrates that when the coding rate $R$ satisfies
	{\small 	$$
		R \leq \frac{1}{2}\log(1 + h^2 \cdot \text{SNR}),
		$$}%
	the decoding error probability $P_e$ doubly-exponentially decays to zero as $n \to \infty$.
	
	Furthermore, from (\ref{sk0-pe-1}), an FBL achievable rate $\mathcal{R}_\text{fd}(N,\varepsilon)$ of the model of Figure~\ref{fig0} \cite{xie} is given by 	
	{\small 	 \begin{equation}\label{Rxfd}
			\mathcal{R}_\text{fd}(N,\varepsilon)\!=\!
			\frac{{N \! -\!1}}{2N}\log \left(1+h^2\text{SNR}\right)\!-\!\frac{1}{2N}\!\log (\frac{L}{12 h^{2}\text{SNR}}),
	\end{equation}}%
	where {\small $L=4\left[Q^{-1}\left(\frac{\varepsilon}{2}\right)\right]^2$}. We can check that when $N\rightarrow \infty$, (\ref{Rxfd}) tends to be $\mathcal{C}_\text{fd}$ of (\ref{zzertong}).
	
	\vspace{-5pt}
	\subsection{Modulo-${d}$ function} \label{M_d}
	
	As shown in \cite{JS11}, the quantized noise generated by the receiver's quantizer can be mathematically modeled by modulo-lattice function (MLF), which is explained below.
	
	A lattice $\wedge_{d}$ in $\mathbb{R}$ spanned by a constant  $d$ is denoted as
	\begin{equation}
		\wedge_{d}=\left\lbrace t=d\cdot a|a\in \mathbb{Z}\right\rbrace.
	\end{equation}
	For $x\in \mathbb{R}$, define its nearest neighbor associated with the lattice $\wedge_{d}$ as
	\vspace{-3pt}
	\begin{equation}
		\mathcal{Q}_{\wedge_{d}}(x)\triangleq\text{arg} \min\limits_{t\in \wedge_{d}}\|x-t\|,
	\end{equation}
	and the modulo-${d}$ function $\mathbb{M}_{{d}}[\cdot]$ is the difference between $\mathcal{Q}_{\wedge_{d}}(x)$ and $x$, namely,
	\begin{equation}\label{sigma_z}
		\mathbb{M}_{{d}}[x] \triangleq \mathcal{Q}_{\wedge_{d}}(x)-x.
	\end{equation}
	The following Proposition \ref{M_d-p} provided by \cite{lattice} gives key properties of $\mathbb{M}_{{d}}[\cdot]$ (which will be used in the remainder of this paper),
	see the detail below.
	\begin{proposition}\label{M_d-p}	
		\begin{itemize}
			\item[(i)] $\mathbb{M}_{d}[x]\in [-\frac{d}{2},\frac{d}{2})$;
			\item[(ii)] The modulo distributive law
			\begin{equation*}
				\mathbb{M}_d[\mathbb{M}_d[x+d_1]+d_2-x]=\mathbb{M}_d[d_1+d_2];
			\end{equation*}
			\item[(iii)]  Let ${V}\sim {\rm Unif}([-\frac{{d}}{2},\frac{{d}}{2}))$, then $\mathbb{M}_d[x+V]$ is uniformly distributed over {\small $[-\frac{{d}}{2},\frac{{d}}{2})$} for any $x \in \mathbb{R}$;	
			\item[(iv)] {\small $\mathbb{E}[(\mathbb{M}_d[x+V])^2]=\frac{d^2}{12}$}.
		\end{itemize}
	\end{proposition}
	
	\begin{figure}[htb]
		\vspace{-10pt}
		\centering
		\includegraphics[scale=0.16]{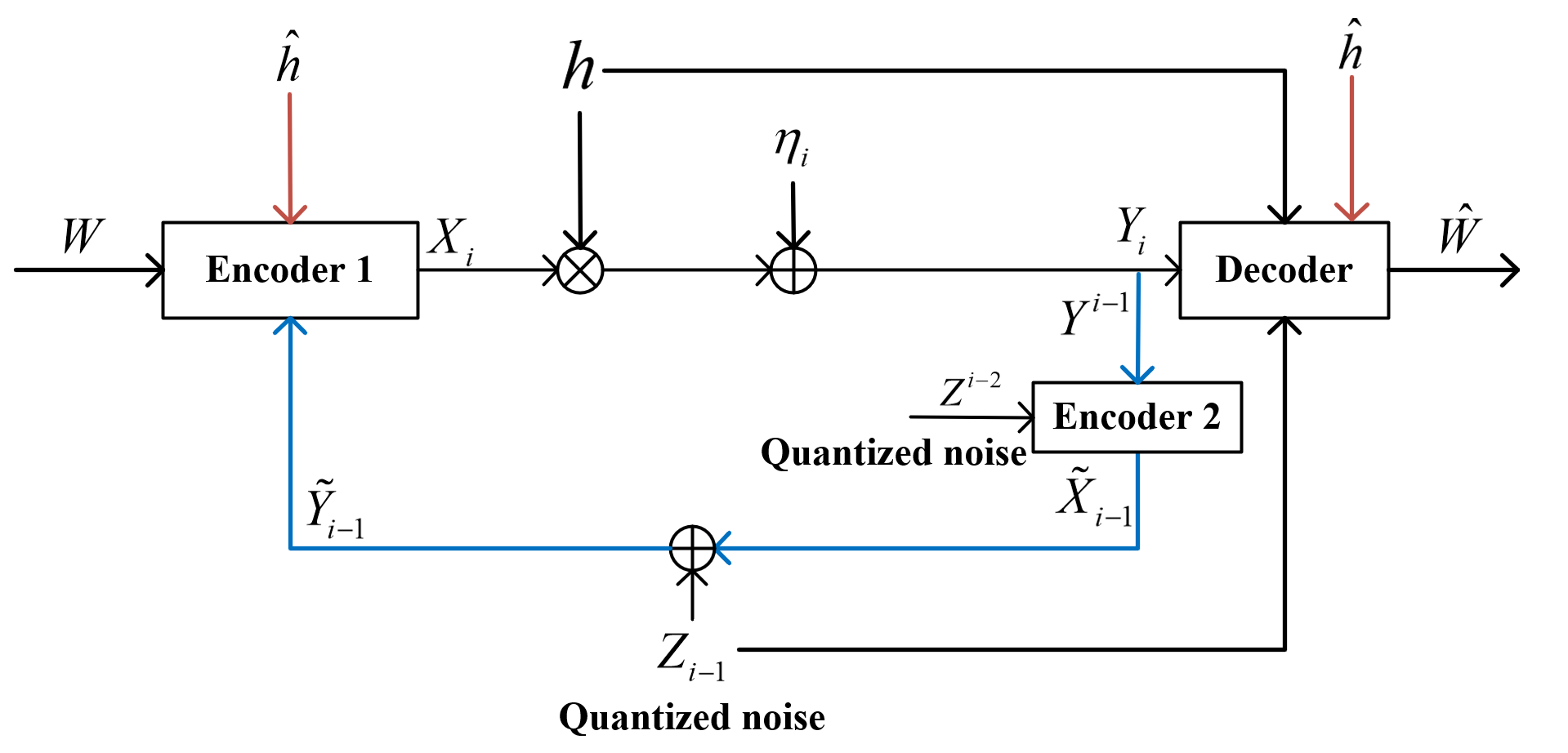}
		\caption{The quasi-static fading channel with I-CSIT and QFC}
		\label{fig1}
		\vspace{-10pt}
	\end{figure}
	
	\section{The Quasi-Static Fading Channel with I-CSIT and QFC}\label{sk1}
	\subsection{System model and main results}\label{model_1}
	In Figure \ref{fig1}, at
	time $i\in\left\lbrace 1,2,\dots,N\right\rbrace $, the channel inputs and outputs are given by
	{\small 	\begin{equation}
			Y_i=hX_i+\eta_i,\,\,\,\,\,\,\,\,\,\,  \widetilde{Y}_{i}=\widetilde{X}_{i}+Z_{i},	
	\end{equation}}%
	where the channel input $X_i$ satisfies average power constraint $\frac{1}{N}\sum\nolimits_{i=1}^{N}\mathbb{E}[{X}_{i}^2]\leq P$,  $h$ is the
	fading coefficient remaining to be a constant within a block \footnote{Such an assumption holds for the scenario with large coherence bandwidth or coherence time, e.g., indoor scenarios with small delay and Doppler spread \cite{H_1}~-\cite{H_2}.}, the AWGN
	$\eta_i\sim\mathcal{N}(0,\sigma^{2})$,
	$\widetilde{X}_i$ is the receiver's feedback codeword with average power constraint
	$\frac{1}{N}\sum\nolimits_{i=1}^{N}\mathbb{E}[\widetilde{X}_{i}^2]\leq \widetilde{P}$, $Y_i$
	and $\widetilde{Y}_i$ are the outputs of the forward and feedback channels, respectively, and
	$Z_i$ is the QFC noise defined below.
	\begin{definition}\label{Z}
		The QFC noise $Z_i$ is generated by the receiver's local quantizer, which is given by
		{\small 	\begin{equation}\label{qfcn}
				Z_i = \mathbb{M}_{2\sigma_z}[\widetilde{X}_{i}],
		\end{equation}}
		where $\sigma_z$ measures the fineness of the quantizer \cite{JS11} and $\sigma_z\geq 0$.
	\end{definition}
	
	\begin{definition}\label{norm}
		Assume that $h$ is perfectly known by the receiver, and the transmitter
		only knows an estimation about $h$, namely, $\hat{h}$. Here $\hat{h}$ is also assumed to be known by the receiver since it is based on the feedback signal. Furthermore, note that the norm-bounded distortion $$\bigtriangleup\triangleq \|h-\hat{h}\|_2$$ between $h$ and $\hat{h}$ has been commonly used in the fading channels with imperfect CSI at the transmitter \cite{i-csi}, hence in this paper we adopt this similar argument to measure the distortion about $h$.
	\end{definition}
	\begin{definition} An $(M, N, P, h, \hat{h}, \bigtriangleup)$-code for the model of Figure \ref{fig1} consists:
		\begin{itemize}
			\item 	The message $W$ is uniformly drawn from $\mathcal{W} = \left\lbrace 1, 2, . . . , M\right\rbrace $.
			\item 	At time $i$ ($i \in\left\lbrace 1, 2, \dots , N\right\rbrace $),
			the codeword $X_i=g_i(W,\boldsymbol{\widetilde{Y}}^{i-1},\hat{h})$, where
			$g_i$ is the feedforward encoding function, and $\boldsymbol{\widetilde{Y}}^{i-1}=[\widetilde{Y}_1,\dots,\widetilde{Y}_{i-1}]^{T} $ is the feedback signal at previous time instants.
			
			\item At time index $i$, the feedback codeword \footnote{Here note that I-CSIT yields additional error in the SK-type encoding-decoding procedure, hence the active feedback is necessary for eliminating this type of error.}	$\widetilde{X}_{i}=\widetilde{g}_i(\boldsymbol{Y}^{i},h,\hat{h},\bigtriangleup,\boldsymbol{Z}^{i-1})$, where $\widetilde{g}_i$ is the feedback encoding function, $\boldsymbol{Y}^{i}=[{Y}_1,\dots,{Y}_{i}]^{T}$ and $\boldsymbol{Z}^{i-1}=[{Z}_1,\dots,{Z}_{i-1}]^{T}$ are the previous time's forward channel output and the feedback QFC noise, respectively.
			
			\item At time $N$, the output of the decoder is  $\hat{W}=\psi(\boldsymbol{Y}^{N},h,\hat{h},\bigtriangleup,\boldsymbol{Z}^{N-1})$, where $\psi$ is the decoding function, and the average decoding error probability is defined as
			\vspace{-3pt}
			{\small 			\begin{equation}\label{pe}
					P_{e}=\frac{1}{M}\sum\limits_{w = 1}^{M}\Pr\{\hat{W}\neq w|
					w\;\mbox{was sent}\}.
					\vspace{-5pt}
			\end{equation}}%
		\end{itemize}
	\end{definition}
	The rate ${R}$ is said to be $(N,\varepsilon,D)$-achievable if for given coding blocklength  $N$, error probability $\varepsilon$  and a targeted estimation distortion $D$ about $h$, there exist encoders and decoders such that
	{\small 	\begin{equation}
			\frac{1}{N}\log M\geq {{R}}-\varepsilon, \,\,\,\, P_{e}\leq \varepsilon, \,\,\,\, \bigtriangleup \leq D.
	\end{equation}}%
	The $(N,\varepsilon,D)$-capacity of the quasi-static fading channel with I-CSIT and QFC is  the
	supremum over all $(N,\varepsilon,D)$-achievable rates defined above, and
	it is denoted by  $\mathcal{C}_\text{I-CSIT}^\text{QFC}(N,\varepsilon,D)$.
	
	\vspace{-8pt}
	\subsection{ Main result}
	\begin{theorem}\label{th1}	
		For given coding blocklength $N$, error probability $\varepsilon$ and targeted estimation distortion $D$ about $h$, the $(N,\varepsilon,D)$-capacity $\mathcal{C}_\text{I-CSIT}^\text{QFC}(N,\varepsilon,D)$ is lower bounded by
		{\small  			\begin{align}
				&\mathcal{C}_\text{I-CSIT}^\text{QFC}(N,\varepsilon,D)\geq\mathcal{R}_\text{I-CSIT}^\text{QFC}(N,\varepsilon,D)\\
				&=\frac{{N-1}}{2N}\log (1+H^2\text{SNR}\cdot\frac{A}{B})
				\!-\!\frac{1}{2N}\log (\frac{L}{12\cdot H ^{2}\text{SNR}}),\nonumber						
		\end{align}}%
		where
		\vspace{-5pt}
		{\small 	\begin{equation}
				\begin{aligned}
					\setlength\abovedisplayskip{0.1cm}
					H&=[\mid \hat{h}\mid-{D}]^{+},\,\,\,\,\,\,\,\, L=4\left[Q^{-1}\left(\frac{\varepsilon}{4}\right)\right]^2, \\
					B&=(\sqrt{A} +\sigma_z )^2+\frac{3\widetilde{P}\cdot\varepsilon}{2},\,\,\,\,\,\text{SNR}=\frac{P}{\sigma^{2}},\nonumber\\	
					A&=(( \sqrt{3\widetilde{P}}-\sigma_z)\cdot[ Q^{-1}( \frac{\varepsilon}{4(N-1)}) ] ^{-1}) ^2.\nonumber	
					\setlength\belowdisplayskip{0.1cm}
				\end{aligned}
		\end{equation}}%
		\begin{IEEEproof}
			See Section \ref{TH1}.
		\end{IEEEproof}
	\end{theorem}	
	
	\begin{remark}\label{R1}
		
		\begin{itemize}
			
			\item \textit{(Upper Bound):} Since {\small $\mathcal{C}_\text{I-CSIT}^\text{QFC}(N,\varepsilon,D)$} is no larger than the rate of the same model with perfect CSI at the transceiver, $\mathcal{R}_\text{fd}(N,\varepsilon)$ of (\ref{Rxfd}) serves as a trivial upper bound on $\mathcal{C}_\text{I-CSIT}^\text{QFC}(N,\varepsilon,D)$.
			
			\item {\textit{(Comparison of $\mathcal{R}_\text{I-CSIT}^\text{QFC}(N,\varepsilon,D)$ and $\mathcal{R}_\text{fd}(N,\varepsilon)$):} The rate $\mathcal{R}_\text{I-CSIT}^\text{QFC}(N,\varepsilon,D)$ cannot completely reduce to that of the same model with perfect CSI and noiseless feedback ($\mathcal{R}_\text{fd}(N,\varepsilon)$ given in (\ref{Rxfd})) when $D \to 0$ and $\sigma_z \to 0$,
				and this imperceptible difference is caused by introducing modulo lattice function and the assumption of active feedback in this paper.}
			
			\item \textit{(Interpretation of $H$):} Theorem \ref{th1} is obtained by an SK-type scheme which will be shown later. Besides this,
			another crucial point to understand Theorem \ref{th1} is the use of $H$, and the intuition behind this is explained below.
			
			Recall that $\mathcal{C}_\text{fd}=\frac{1}{2}\log(1+h^{2}\cdot\text{SNR})$ is the channel capacity of the quasi-static fading channel with perfect CSI at the transceiver and noiseless feedback. However, the transmitter does not know the real value of $h$,
			hence he cannot compute this capacity. Instead, since the transmitter knows $\hat{h}$ and the targeted estimation distortion $D$, then it is natural to ask: from the transmitter's point of view, what is the reliable transmission rate below $\mathcal{C}_\text{fd}$ he should adopt for successful communication?
			
			To answer this question, a lower bound on $h^2$ should be established by the transmitter. First, note that if
			$\hat{h} \geq D$, from $|\hat{h}-h|\leq D$, it is easy to check that $ (\hat{h}-D)^{2} \leq h^{2}$, which indicates that $(\hat{h}-D)^{2}$ serves as a lower bound on $h^2$, and
			the corresponding achievable rate $\frac{1}{2}\log(1+(\hat{h}-D)^{2}\cdot\text{SNR})$ is known and can be adopted by the transmitter.
			
			Analogously, if $\hat{h} \leq -D$, we have $ (\hat{h}+D)^{2} \leq h^{2}$, and
			$\frac{1}{2}\log(1+(\hat{h}+D)^{2}\cdot\text{SNR})$ can be adopted by the transmitter for
			reliable transmission.
			
			Finally, note that if $-D < \hat{h} < D$, we can only get a trivial lower bound $h^{2}\geq 0$,
			which implies that in such a case, no positive reliable transmission rate can be adopted by the transmitter.
			
			Hence in Theorem \ref{th1}, we adopt $H=[\mid \hat{h}\mid-{D}]^{+}$ covering all above three cases
			to characterize the actual rate of the SK-type scheme, and it is easy to check that
			{\small \begin{equation}\label{H-h}
					H^{2}\leq h^{2}.
			\end{equation}}
			
			\item {\textit{(Coding complexity):} In our scheme of Theorem \ref{th1}, there are two kinds of operations, where one is the MMSE based linear coding operation, and the other is the modulo lattice coding operation. Since the complexity of linear coding is $\mathcal{O}(N)$, and that of modulo lattice coding is $\mathcal{O}(N\log N)$, the overall complexity of our scheme is $\mathcal{O}(N\log N)$.}
			
		\end{itemize}

	\end{remark}

	\begin{figure}[htb]
		\vspace{-10pt}
		\centering
		\includegraphics[scale=0.19]{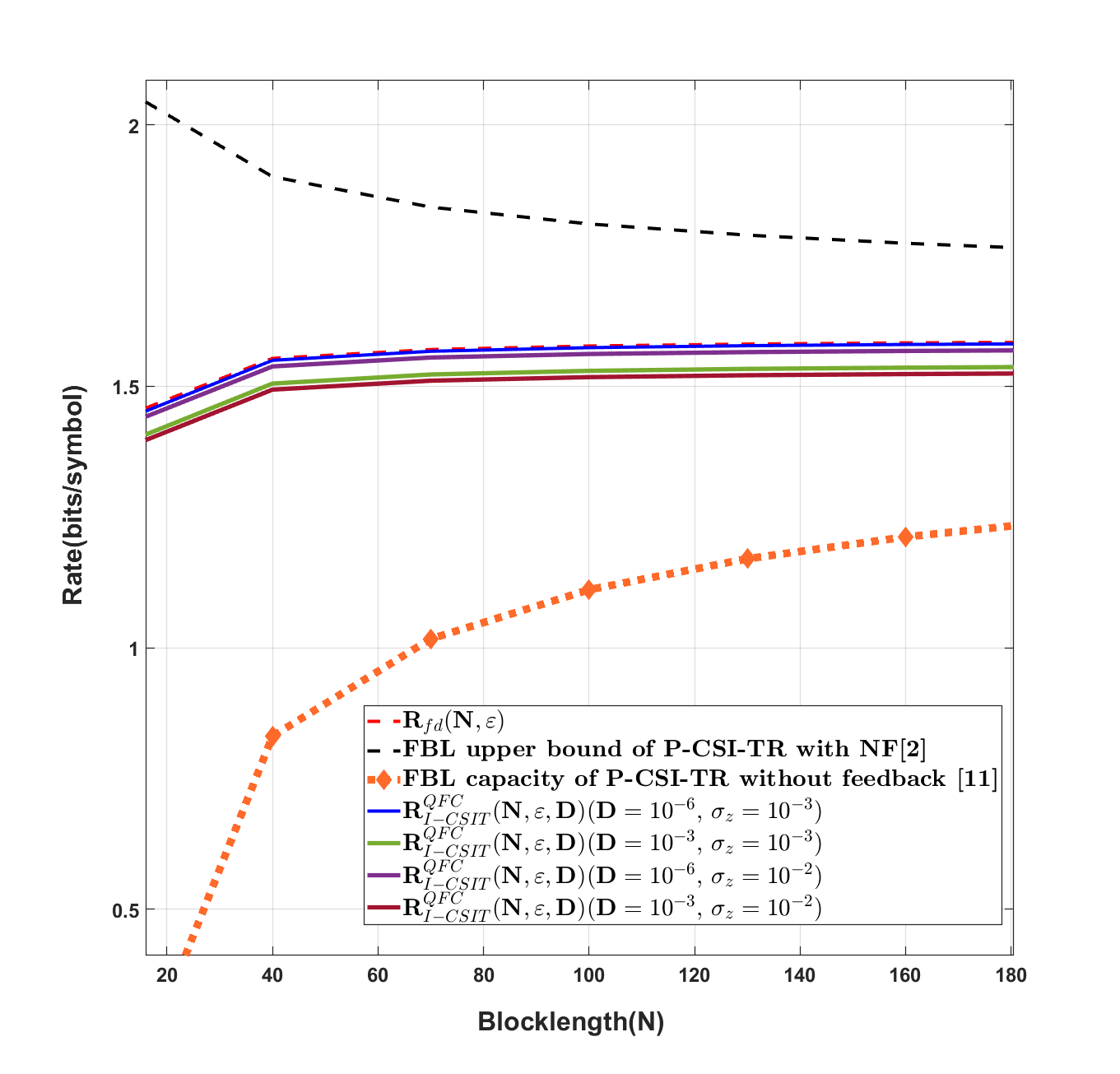}
		\caption{{\small {Rate versus blocklength $N$ for $\text{SNR} =10$, $\varepsilon=10^{-6}$, $h=0.9$ and  $\widetilde{P}=10$}}}
		\label{fg2}
		\vspace{-10pt}
	\end{figure}		
	
	\subsection{Numerical  results}	
	
\begin{figure}[htb]
		\centering
		\includegraphics[scale=0.2]{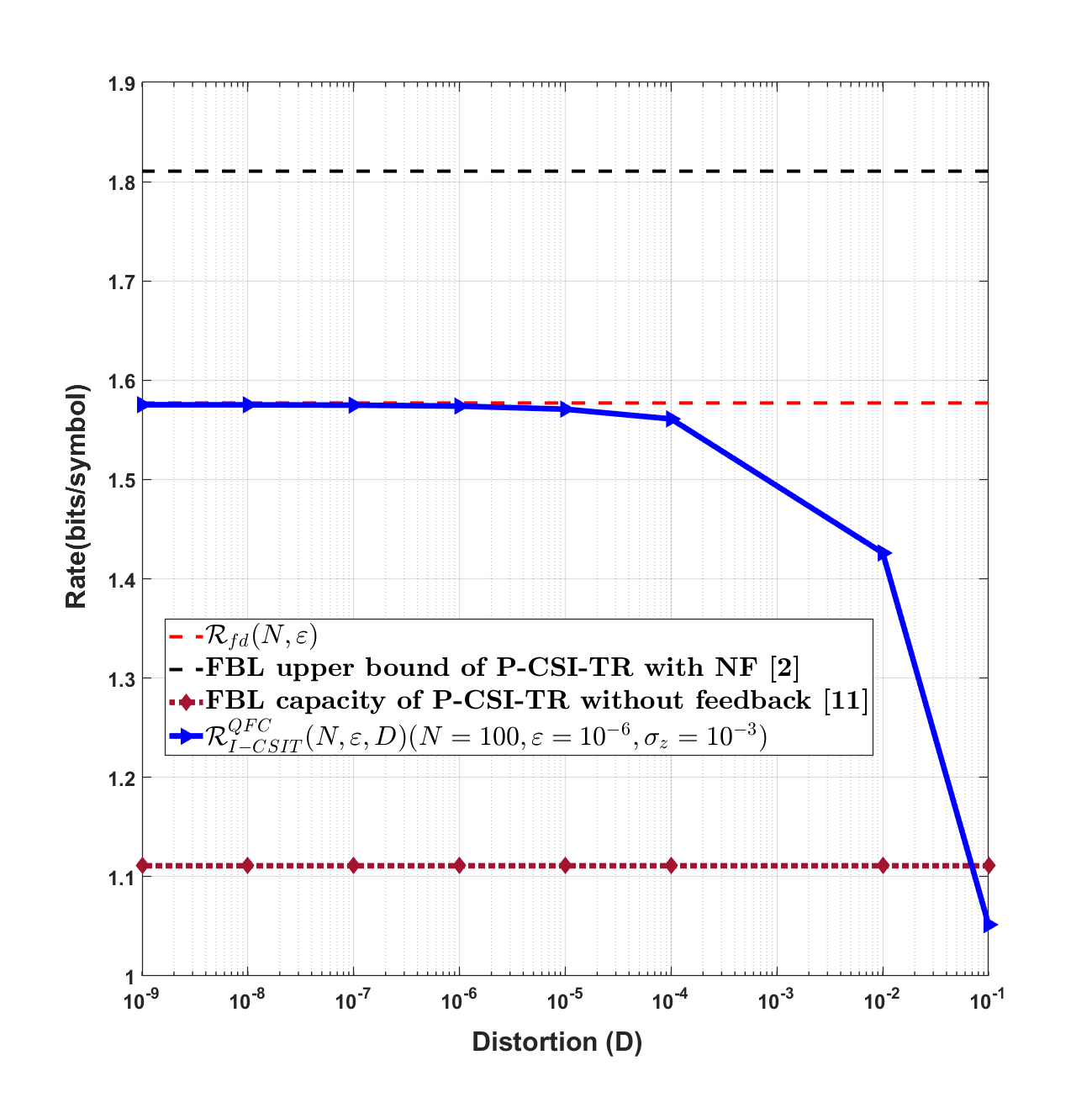}
		\caption{{\small { Rate versus distortion $D$ for $N =100$, $\text{SNR} =10$, $\sigma_z=10^{-3}$, $\varepsilon=10^{-6}$, $h =0.9$ and  $\widetilde{P}=10$}}}
		\label{fg3}
	\end{figure}

	Here note that when the transmitter's estimation distortion about the CSI and the quantized noise of the feedback channel tend to zero, our model studied in this section reduces to  		
	the channel with perfect CSI at the transceiver and noiseless feedback, and Figures \ref{fg2} and \ref{fg3} show that our results coincide with the above facts.
	
	Besides this, it is well known that I-CSIT causes performance loss in the fading channels. However,
	from Figures \ref{fg2} and \ref{fg3} we see that through quantized feedback,
	we can construct effective coding scheme which still brings rate gain to the quasi-static fading channel with I-CSIT. Moreover, similar to the property of the FBL coding, Figure~\ref{fg2} shows
	that our rate is increasing and approaching its Shannon limit when the coding blocklength $N$ is increasing.
	
	{The SK-type scheme of \cite{JS11} is used as a benchmark scheme. Under the same experiment setting, the following Figure \ref{17vs} shows that our scheme outperforms the benchmark scheme.}

	\begin{figure}[htb]
		\centering
		\includegraphics[scale=0.19]{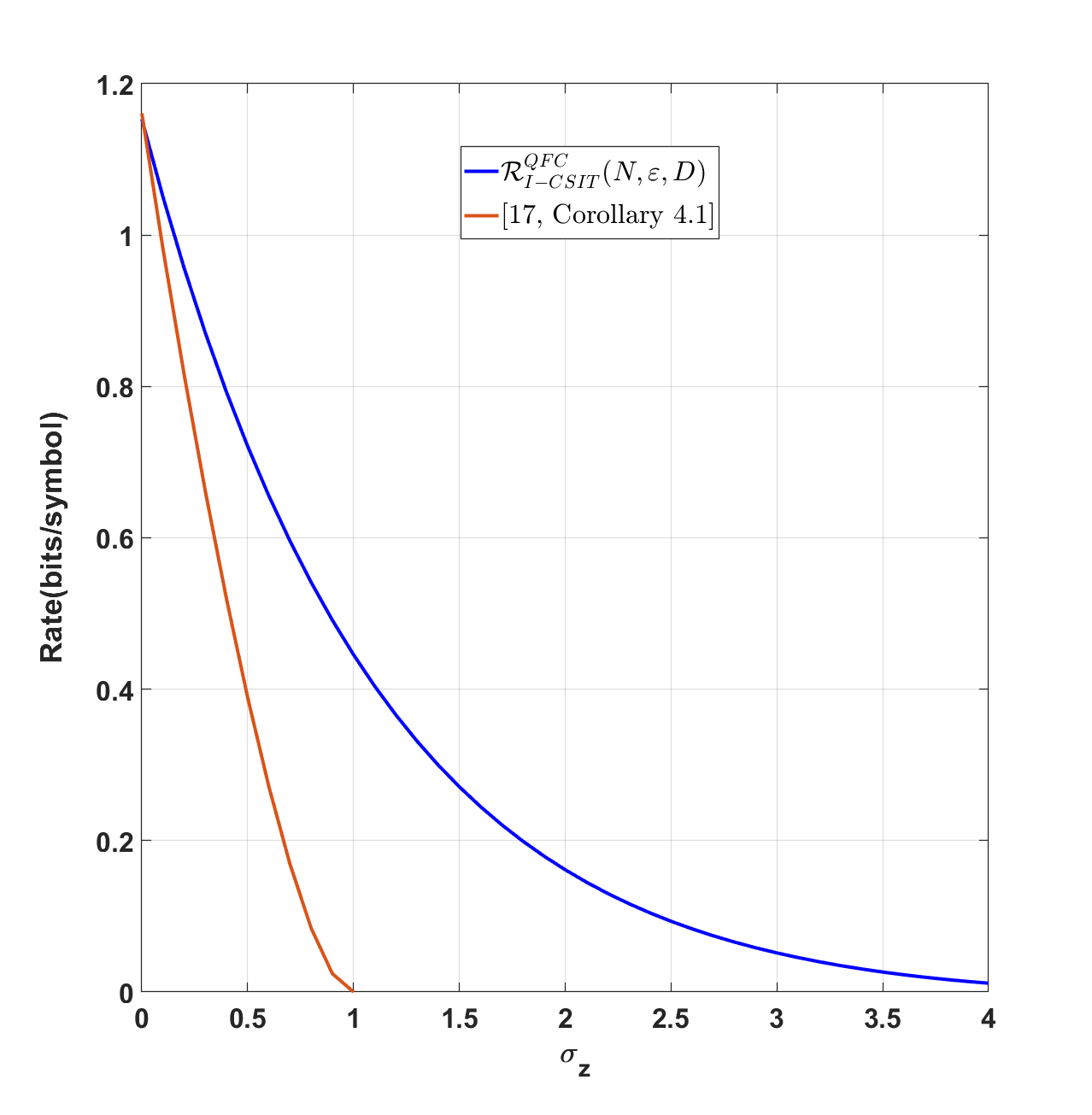}
		\caption{{\small { Rate versus $\sigma_z$ for $\text{SNR} =4$, $h =1$, $D=0$, $\widetilde{P}=10$,  $N=200$  and $\varepsilon=10^{-4}$}}}
		\label{17vs}
	\end{figure}

	
	\vspace{-10pt}
	\subsection{ Proof of Theorem \ref{th1}}\label{TH1}

	In this subsection, we show the detail of our scheme and provide a corresponding formal performance analysis of it.

	
	
	\subsubsection{Encoding-decoding procedure of our scheme} \label{sk-s1}
	
	Similar to Section \ref{sk0}, we map the message $W$ to a real value $\theta$ which is the midpoint of the message's sub-interval.
	
	$\textbf{Initialization:}$ At time instant $1$, the transmitter encodes $\theta$ as $X_{1}=\sqrt{{12P}}\theta$.
	
	Once receiving $Y_{1}=hX_{1}+\eta_{1}$, the receiver computes his first estimation $\hat{\theta}_{1}$ of $\theta$ by
	{\small 		\begin{equation}
			\hat{\theta}_{1}=\frac{Y_{1}}{h\sqrt{12P}}=\theta+\frac{\eta_{1}}{h\sqrt{12P}},
	\end{equation}}%
	then the receiver sends
	{\small 		\begin{equation}
			\widetilde{X}_{1}=\mathbb{M}_{\widetilde{d}}[\gamma_{1}\cdot\hat{\theta}_{1}+V_1],
	\end{equation}}%
	where $\widetilde{d}=\sqrt{12\widetilde{P}}$,
	$V_{1}$ is a dither signal \footnote{{The dither signal is commonly used in the modulo-lattice structure, and to the best of the authors' knowledge, most of the existing works in the field of lattice coding assume that the dither signal is known to all parties. One practical way to realize
			this assumption is by the leverage of pseudo random number generator, just like that used in TS 38.211 protocol \cite{3GPP}.} } uniformly distributed in $[-\frac{\widetilde{d}}{2},\frac{\widetilde{d}}{2})$, it is independent of all other random variables and known by all parties, and $\gamma_{1}$ will be defined later.
	
	The feedback signal received by the transmitter is denote by $\widetilde{Y}_{1}=\widetilde{X}_{1}+Z_{{1}}$.
	
	$\textbf{Iteration:}$	
	At time instant $i \in \left\lbrace {2},\ldots,N\right\rbrace$, after the transmitter receives the feedback $\widetilde{Y}_{i-1}$, he sends
	{\small     	\begin{align}\label{X1}
			X_{i}&=\alpha\cdot \mathbb{M}_{\widetilde{d}}[\widetilde{Y}_{i-1}-\gamma_{i-1}\cdot\theta-V_{i-1}]\\
			&\mathop{=}\limits^{\text{(a)}}\alpha\cdot\mathbb{M}_{\widetilde{d}}[\gamma_{i-1}\cdot\epsilon_{i-1}+Z_{i-1}],\nonumber
	\end{align}}%
	where $\text{(a)}$ follows from property (ii) of Proposition \ref{M_d-p}, $V_{i-1}$
	is a dither signal uniformly distributed in $[-\frac{\widetilde{d}}{2},\frac{\widetilde{d}}{2})$, and it is independent of all other random variables and known by all parties, and $\alpha$, $\gamma_{i-1}$ will be defined later.
	
	Once receiving
	{\small 		\begin{equation}\label{sk1-yi}
			Y_{i}=hX_{i}+\eta_{i},
	\end{equation}}%
	the receiver calculates the auxiliary signal
	{\small 		\begin{equation}\label{aux_sk1}
			\dot{Y}_{i}=Y_{i}-h\alpha\cdot Z_{i-1},
	\end{equation}}%
	and updates his estimation  $\hat{\theta}_{i}$ of $\theta$ by
	{\small 		\begin{equation}\label{sk1-beta}
			\hat{\theta}_{i}=\hat{\theta}_{i-1}-\beta_{i-1}\dot{Y}_{i}.
	\end{equation}}%
	where $\beta_{i-1}$  will be defined in (\ref{sk1-beta-p}). Define $\epsilon_{i}=\hat{\theta}_{i}-\theta$, then (\ref{sk1-beta}) yields that
	{\small 		\begin{equation}\label{sk1-ep-r}
			\epsilon_{i}=\epsilon_{i-1}-\beta_{i-1}\dot{Y}_{i}.
	\end{equation}}%
	
	The receiver sends $\widetilde{X}_i$ back to the transmitter via the QFC, where
	{\small 		\begin{equation}
			\widetilde{X}_i=\mathbb{M}_{\widetilde{d}}[\gamma_{i}\cdot\hat{\theta}_{i}+V_{i}].
	\end{equation}}%
	
	$\textbf{Decoding:}$
	The receiver's final decoding
	process is the same as that of Section \ref{sk0}, and we omit the description here.
	
	\subsubsection{ Performance analysis}\label{sk1-p}
	
	Here note that the decoding error probability analysis of the above scheme is difficult due to the involving of modulo-lattice functions. To deal with this problem, along the lines of \cite{JS10},
	we adopt a coupled system where all modulo-lattice functions are removed and
	some other parts of the original system are changed,
	and the detailed explanation of the coupled system is given below.
	\begin{itemize}
		\item All modulo-lattice functions in the original system are removed in the coupled system.
		\item The channel coefficient ($h$) and the noises ($\eta_i$ and $Z_i$) in the coupled system are the same as those in the original system.
		\item All parameters $\alpha$, $\beta_i$, and $\gamma_i$ in the original system are preserved in the coupled system.
		\item In the coupled system, except the above unchanged parts of the original system,
		other signals and all error events of the original system are changed and marked by adding
		prime symbol ($'$) to original ones.
	\end{itemize}
	In the remainder of this subsection, first, we analyze the error probability of the coupled system, and next we show that it is larger than the decoding error probability of the original system. Finally, we show the formal performance analysis of the original system.
	
	For $i \in \left\lbrace {1},\ldots,N-1\right\rbrace$, define $E_i$ as the event where a modulo-aliasing error occurs in original system, and $E_i$ is given by
	\vspace{-1pt}
	{\small 		\begin{equation}\label{Ei_1}
			E_i\triangleq \lbrace \gamma_{i}\cdot\epsilon_{i}+Z_{i}\notin [ -\frac{\widetilde{d}}{2},\frac{\widetilde{d}}{2}) \rbrace.
	\end{equation}}%
	Furthermore, we define $E_{N}$ as the receiver's decoding error event in original system, and it is given by
	{\small 		\begin{equation}\label{Ei_N}
			E_{N}\triangleq\lbrace \epsilon_{N} \notin [ -\frac{1}{2^{NR+1}},\frac{1}{2^{NR+1}}) \rbrace.
			\vspace{-3pt}
	\end{equation}}%

Now we turn to the coupled system,
let $E_i^{'} \triangleq \lbrace \gamma_{i}\cdot\epsilon'_{i}+Z_{i} \notin [-\frac{\widetilde{d}}{2},\frac{\widetilde{d}}{2})\rbrace$ and   $E_{N}^{'}\triangleq\lbrace \epsilon'_{N} \notin [ -\frac{1}{2^{NR+1}},\frac{1}{2^{NR+1}}) \rbrace$.

Here note that for the decoding error probability $P_e$ in the original system, the key step to establish its upper bound is to show that
{\small \begin{equation}\label{sk1-pe-1}
		P_e\leq\sum_{j=1}^N\Pr\left\{E_j'\right\},
\end{equation}}%
which follows from
{\small \begin{equation}\label{model1-1}
		\Pr\{\bigcup_{j=1}^{N}E_j \}= \Pr\{\bigcup_{j=1}^{N}E_j' \},
\end{equation}}%
and {\small $P_e\leq\Pr\{\bigcup_{j=1}^{N}E_j\}$}, where the proof of (\ref{model1-1}) is 
similar to that of \cite{JS10}, and we omit it here.

Then to further bound $P_e$, we need to calculate the single error event {\small $\Pr\{E_i'\}$} ({\small $i\in\{1,\cdots,N-1\}$}) and {\small $\Pr\{E_N'\}$ }of the coupled system. First, we compute {\small $\Pr\{E_i'\}$ ($i\in\{1,\cdots,N-1\}$)}, which is denoted by
{\small  \begin{align}\label{sk1-gamma1-1}	
		\Pr\left\lbrace E_i^{'}\right\rbrace  &\mathop{=}\limits^{\text{(b)}}\Pr\left\{\arrowvert\gamma_i{\epsilon}_i^{'}+Z_i\arrowvert\geq\sqrt{3\widetilde{P}}\right\}\nonumber\\
		&\mathop{\leq}\limits^{\text{(c)}} \Pr\left\{\arrowvert\gamma_i{\epsilon}_i^{'}\arrowvert+\sigma_z\geq\sqrt{3\widetilde{P}}\right\},						
\end{align}}%
where $\text{(b)}$ follows from {\small $\widetilde{d}=\sqrt{12\widetilde{P}}$}, and
$\text{(c)}$ follows from $Z_i = \mathbb{M}_{2\sigma_z}[\widetilde{X}_{i}]$ and $|Z_i|\leq \sigma_z$.
To further bound $\Pr\{E_i'\}$ of (\ref{sk1-gamma1-1}), we need to determine
${\epsilon}_i^{'}$ and
{\small \begin{equation}\label{sk1-alpha}	
		\alpha_i' = \mathbb{E}[\epsilon_i^{'2}]
\end{equation}}%
of the coupled system first, where ${\epsilon}_i^{'}$ is given by
{\small \begin{equation}\label{sk1-ep1}	
		\epsilon_{i}' = \epsilon_{i-1}' - \beta_{i-1} \cdot \dot{Y}_{i}'.
\end{equation}}%
Here (\ref{sk1-ep1}) is in analogous to (\ref{sk1-ep-r}) of the original system,
and note that the all parameters in the original system are preserved in the coupled system.
To determine ${\epsilon}_i^{'}$ and $\alpha_i'$, it remains to
calculate the terms $\dot{Y}_{i}'$ and $\beta_{i-1}$ of (\ref{sk1-ep1}), see the detail below.

\emph{Calculation of $\dot{Y}_{i}'$}:
First, recall that in the original system, based on \eqref{X1} and \eqref{sk1-yi}, we have
{\small \begin{equation}
		Y_i=h\alpha\cdot\mathbb{M}_{\widetilde{d}}[\gamma_{i-1}\cdot\epsilon_{i-1}+Z_{i-1}]+\eta_{i},
\end{equation}}%
and the auxiliary signal $\dot{Y}_i$ of (\ref{aux_sk1}) is re-written by
{\small \begin{equation}\label{sk1-dotY}
		\dot{Y}_i=h\alpha\cdot\mathbb{M}_{\widetilde{d}}[\gamma_{i-1}\cdot\epsilon_{i-1}+Z_{i-1}]+\eta_{i}-h\alpha\cdot Z_{i-1}.
\end{equation}}%
Next, note that all modulo-lattice functions in the original system are removed in the coupled system, hence we have
{\small \begin{align}\label{eq:Ydot}
		\dot{Y}_i' &=h\alpha\cdot(\gamma_{i-1}\cdot\epsilon_{i-1}'+Z_{i-1})+\eta_{i}-h\alpha\cdot Z_{i-1}\\
		&= h\alpha\gamma_{i-1}\cdot\epsilon_{i-1}' + \eta_i,\nonumber	
\end{align}}%
which implies that the QFC noise $Z_i$ in the coupled system can be perfectly eliminated by using the auxiliary signal $\dot{Y}_i'$.

\emph{Calculation of $\beta_{i-1}$}: Note that $\beta_{i-1}$ is MMSE coefficient for both original and coupled systems, and it is given by
{\small \begin{equation}\label{sk1-beta-p}
		\beta_{i-1} = \frac{\mathbb{E}[\epsilon_{i-1}' \cdot \dot{Y}_{i}']}{\mathbb{E}[(\dot{Y}_{i}')^2]}.
\end{equation}}%
Since $\beta_{i-1}$ is correlated with $\epsilon_{i-1}'$, currently it cannot be calculated, and we will determine it later.

Based on the determination of $\dot{Y}_{i}'$ and $\beta_{i-1}$, $\alpha_i'$ of (\ref{sk1-alpha}) can be further re-written as
{\small \begin{equation}\label{alpha_i-1}
		\alpha_{1}'=\frac{\sigma^2}{12 Ph^2},\qquad
		\alpha_i'
		=(1+\frac{h^{2}\alpha^2\gamma_{i-1}^2\alpha_{i-1}'}{\sigma^2})^{-1}\alpha_{i-1}'.
\end{equation}}%

From (\ref{eq:Ydot}), we conclude that $\dot{Y}_i'$ is Gaussian distributed, leading to $\epsilon_{i}^{'}$ of (\ref{sk1-ep1}) also be Gaussian distributed. Hence
(\ref{sk1-gamma1-1}) can be further re-written as
{\footnotesize  \begin{equation}\label{sk1-gamma-2}	
		\Pr\left\lbrace E_i^{'}\right\rbrace \!\leq \! \Pr\left\{\arrowvert\gamma_i{\epsilon}_i^{'}\arrowvert+\sigma_z \!\geq\!\sqrt{3\widetilde{P}}\right\}
		=2Q(\frac{\sqrt{3\widetilde{P}}-\sigma_z }{ \sqrt{\gamma_i^2\alpha_i'}}).								
\end{equation}}
Though (\ref{sk1-gamma-2}) provides a upper bound on $\Pr\lbrace E_i^{'}\rbrace$,
$\alpha_i'$ given in (\ref{alpha_i-1}) is not known by the transmitter since $h$ is not available at him. Based on the choice of $H$
explained in Remark \ref{R1}, define
{\footnotesize  \begin{equation} \label{Hsigma}
		\alpha_{1}'(H) = \frac{\sigma^2}{12 P H^2},\quad
		\alpha_{l+1}'(H) = (\frac{1}{\alpha_{l}'(H)} + \frac{H^{2}  \alpha^2  \gamma_l^2}{\sigma^2})^{-1}  ,		
\end{equation}}%
where $l\in\{1,2,...,N-1\}$ and {\small $H = \max(|\hat{h}| - D, 0)$}.
Next, since
{\small \begin{equation}\label{sk1-alpha_j}
		\alpha_{j}'(H) \geq \alpha_{j}'
\end{equation}}%
for $j\in\{1,2,...,N\}$ and $Q(x)$-function monotonically decreases as $x$ increases, (\ref{sk1-gamma-2}) can be further upper bounded by
{\small \begin{equation}\label{sk1-gamma-3}	
		\Pr\lbrace E_i^{'}\rbrace\leq 2Q(\frac{\sqrt{3\widetilde{P}}-\sigma_z }{ \sqrt{\gamma_i^2\cdot\alpha_i'}})
		\leq 2Q(\frac{\sqrt{3\widetilde{P}} - \sigma_z}{\sqrt{\gamma_i^2 \cdot \alpha_{i}'(H)}} ).								
\end{equation}}%
Substituting (\ref{sk1-gamma-3}) into (\ref{sk1-pe-1}), we get
{\small \begin{align}\label{sk1-pe3}
		P_e&\leq\sum_{i=1}^{N-1}\Pr\{E_i'\}+\Pr\{E_N'\}\nonumber\\
		&\leq\sum_{i=1}^{N-1}2Q(\frac{\sqrt{3\widetilde{P}} - \sigma_z}{\sqrt{\gamma_i^2 \cdot \alpha_{i}'(H)}} )+	\Pr\{E_N'\}.				
\end{align}}%
Now it remains to deal with $\gamma_i$, $\alpha$ and $\Pr\{E_N'\}$ of (\ref{sk1-pe3}), see the detail below.

\emph{An upper bound on $\Pr\{E_N'\}$ of (\ref{sk1-pe3})}: Note that
{\small \begin{align}\label{sk1-Pe1}
		&\Pr\{E_{N}'\}=\Pr\lbrace \epsilon_{N}' \notin [ -\frac{1}{2^{NR+1}},\frac{1}{2^{NR+1}}) \rbrace \nonumber\\
		&=2\mathop{Q}( \frac{1}{2^{NR+1}\sqrt{\alpha_{N}'}}) 	
		\mathop{\leq}\limits^{\text{(d)}} 2\mathop{Q}( \frac{1}{2^{NR+1}\sqrt{\alpha_{N}'(H)}}),	
\end{align}}%
where $\alpha_N' = \mathbb{E}[\epsilon_N^{'2}]$ and (d) follows from (\ref{sk1-alpha_j}).

\emph{Determination of $\gamma_i$ and $\alpha$}:
To ensure $P_e$ of (\ref{sk1-pe3}) satisfying $P_e \leq \varepsilon$, we set
{\small \begin{equation}\label{sk1-up0}
		\Pr\left\lbrace E_{N}^{'}\right\rbrace\leq 2\mathop{Q}( \frac{1}{2^{NR+1}\sqrt{\alpha_{N}'(H)}})=\frac{\varepsilon}{2}
\end{equation}}%
and
{\small \begin{equation}\label{sk1-up1}
		2Q(\frac{\sqrt{3\widetilde{P}} - \sigma_z}{\sqrt{\gamma_i^2 \cdot \alpha_{i}'(H)}})	=p'_m=\frac{\varepsilon}{2(N-1)}
\end{equation}}%
for $i\in\{1,\cdots,N-1\}$, which indicates that
{\small \begin{equation}\label{sk1-up2}
		\sum_{i=1}^{N-1}2Q(\frac{\sqrt{3\widetilde{P}} - \sigma_z}{\sqrt{\gamma_i^2 \cdot \alpha_{i}'(H)}} )=\frac{\varepsilon}{2}.
\end{equation}}%
Then from (\ref{sk1-up1}), we have
{\small \begin{equation}\label{sk1-gamma_i}
		\gamma_i = \sqrt{\frac{A}{\alpha_{i}'(H)}},
\end{equation}}%
where {\small $A=(( \sqrt{3\widetilde{P}}-\sigma_z)\cdot[Q^{-1}(\frac{p'_m}{2})]^{-1}) ^2$}.

On the other hand, note that $\alpha$ is a parameter used to guarantee the transmitting power of the original system (see (\ref{X1})), then from (\ref{sk1-gamma_i})
and the power constraint,
it is not difficult to show that if we set
{\small 	\begin{equation}\label{sk1-alpha2}
		\alpha=\sqrt{\frac{P}{B}},
\end{equation}}%
where {\small $B=(\sqrt{A} +\sigma_z )^2+\frac{3\widetilde{P}\cdot\varepsilon}{2}$}, the power constraint
{\small $\mathbb{E}[(X_{i+1})^2]\leq P$}
is ensured.

To this end, it remains to determine $\gamma_i$, which depends on $\alpha_{i}'(H)$.
Interestingly, we find that substituting (\ref{sk1-alpha2}) and (\ref{sk1-gamma_i})
into (\ref{Hsigma}), the general term of {\small $\alpha_{j}'(H)$ ($j\in\{1,2,...,N\}$}) is given by
{\small \begin{equation}\label{sk1-alpha1}
		\alpha_{j}'(H)	=\frac{1}{12\cdot{H}^2\cdot\text{SNR}}\cdot( \frac{1}{1+{H}^2\cdot\text{SNR}\cdot\frac{A}{B}}) ^{j-1}.
\end{equation}}%
Now substituting (\ref{sk1-alpha1}) into (\ref{sk1-gamma_i}),
$\gamma_i$ is determined.

Analogously, substituting (\ref{sk1-alpha1}) into (\ref{sk1-up0}), the rate $R$ can be re-written by
{\small \begin{equation}\label{nitamaqusi-2}
		{R}=
		\frac{1}{2N}\log ( \frac{12\cdot H^{2}\cdot\text{SNR}(1+H^2\cdot\text{SNR}\cdot\frac{A}{B})^{N-1}}{L}),
\end{equation}}%
where {\small  $L=4\left[Q^{-1}\left(\frac{\varepsilon}{4}\right)\right]^2$}.

In addition, substituting (\ref{sk1-ep1}) and (\ref{eq:Ydot}) into (\ref{sk1-beta-p}), and applying
(\ref{sk1-alpha2}) and (\ref{sk1-gamma_i}) to (\ref{sk1-beta-p}), the MMSE coefficient $\beta_{i}$
can be determined. The proof of Theorem \ref{th1} is completed.

\section{The Two-path Quasi-static Fading Channel with I-CSIT and QFC}\label{sk2}
The two-ray channel, which is one of the simplest small-scale fading models in communication systems, has been extensively studied in the literature \cite{ISI5,ISI6,ISI7}. As shown in \cite{memory}, the two-ray channel can be simplified and modeled  as the two-path quasi-static fading channel (TP-QS-FC), and in this section we aim to extend the SK-type scheme to the TP-QS-FC with I-CSIT and QFC.

\subsection{Model formulation and main results}\label{model_2}

\begin{figure}[htb]
	\centering
	\includegraphics[scale=0.19]{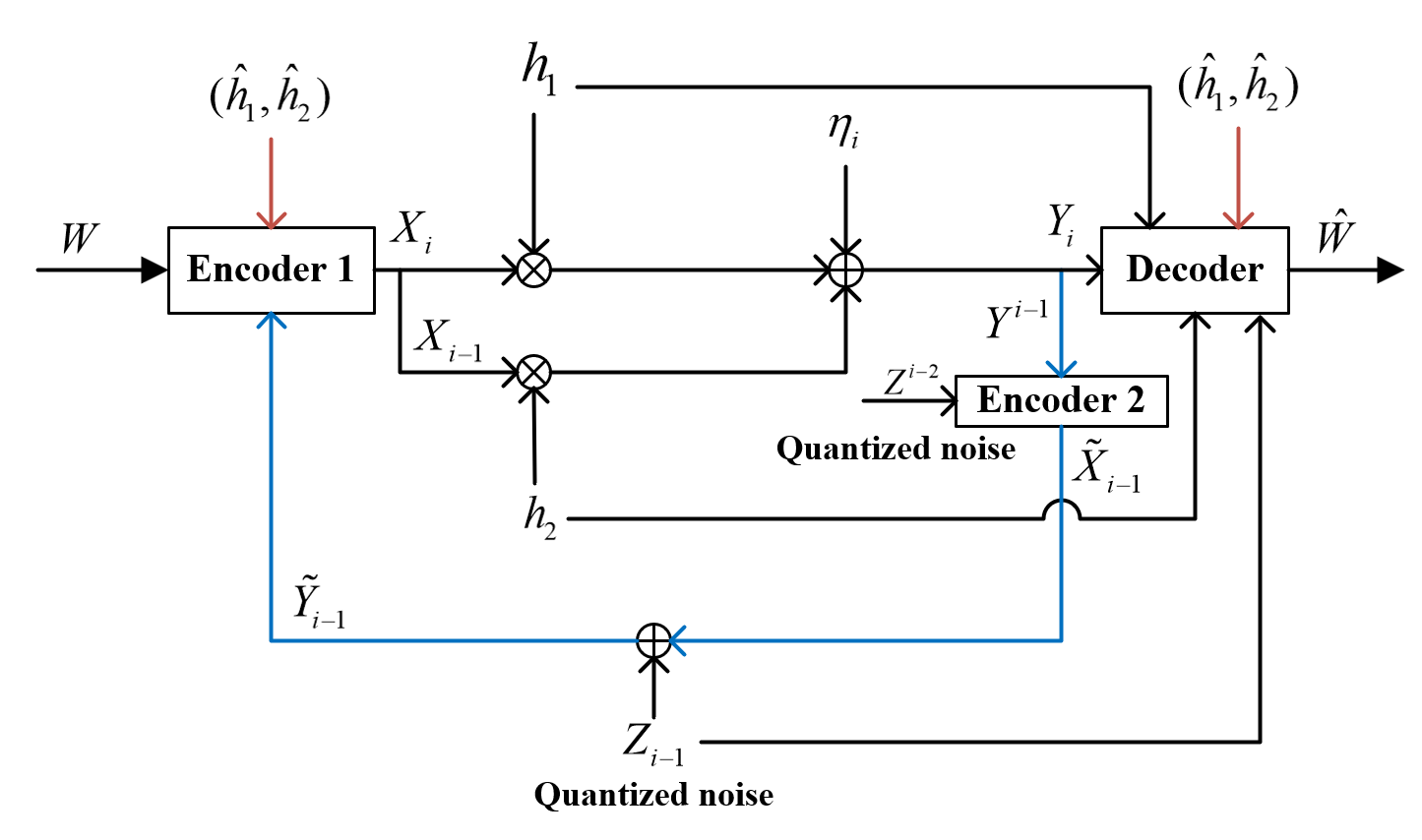}
	\caption{The TP-QS-FC with I-CSIT and QFC}
	\label{fig2}
\end{figure}

The TP-QS-FC with I-CSIT and QFC is depicted in  Figure \ref{fig2}. At time index {\small $i\in\left\lbrace 1,2,\dots,N\right\rbrace $},  the channel inputs and outputs are given by
{\small \begin{equation}
		Y_i=h_{1}\cdot X_i+h_{2}\cdot X_{i-1}+\eta_i,\,\,\,\,\,\,\,\,\, \widetilde{Y}_i=\widetilde{X}_i+Z_i,	
\end{equation}}%
where $X_i$ denotes the channel input at time $i$, subject to an average power constraint  {\small $\frac{1}{N}\sum\nolimits_{i=1}^{N}\mathbb{E}[{X}_{i}^2]\leq P$}, $h_1$ and $h_2$ are two non-zero fading coefficients remaining constants within a block,
the AWGN $\eta_i\sim\mathcal{N}(0,\sigma^{2})$,
$\widetilde{X}_i$ is the feedback channel codeword with average power constraint
$\frac{1}{N}\sum\nolimits_{i=1}^{N}\mathbb{E}[\widetilde{X}_{i}^2]\leq  \widetilde{P}$, $Y_i$
and $\widetilde{Y}_i$ are the outputs of the forward and feedback channels, respectively, and the definition of the quantized noise $Z_i$ is the same as that of Definition \ref{Z}.
\begin{definition}\label{bb1}
	In this model,  $h_1$ and $h_2$ are perfectly known by the receiver, and the transmitter only obtains estimations about them, which are denoted as $\hat{h}_1$ and $\hat{h}_2$, respectively, and assume that the receiver also knows $\hat{h}_1$ and $\hat{h}_2$ since they are estimated by the feedback signals. Similar to Definition \ref{norm}, norm-bounded distortion is adopted, where	
	{\small 	\begin{equation}
			\bigtriangleup_1\triangleq\|h_1-\hat{h}_1\|_2,\,\,\,\,\,\,\,
			\, \bigtriangleup_2\triangleq\|h_2-\hat{h}_2\|_2. 		
	\end{equation} }%
\end{definition}
\begin{definition}	
	Though the transmitter does not know the real values of  $h_1$ and  $h_2$, at the beginning of each transmission block, the function {\small $\sgn{h_1h_2}$} ({\small $\sgn{x} =
		\begin{cases}
			+1, & x \geq 0 \\
			-1, & x < 0
		\end{cases}$}) of  $h_1$ and $h_2$ can be known by the transmitter, and this can be achieved as follows.
	
	In each block, before the transmission of the message, the receiver sends a pilot signal $\widetilde{X}_0$ to the transmitter, where
	{\small 								\begin{equation}\label{X_0}
			\widetilde{X}_0 =
			\begin{cases}
				+2\sigma_z, & \mathrm{sgn}(h_1h_2) = +1, \\
				-2\sigma_z, & \mathrm{sgn}(h_1h_2) = -1.
			\end{cases}
	\end{equation}}%
	The feedback signal received by the transmitter is
	{\small \begin{equation}\label{Z_0}
			\widetilde{Y}_0 = \widetilde{X}_0 + Z_0 = \widetilde{X}_0 + \mathbb{M}_{{2\sigma_z}}[\widetilde{X}_0]= \widetilde{X}_0,
	\end{equation}}%
	where $Z_0$ is the quantized noise during the transmission of $\widetilde{X}_0$ in QFC,
	and according to Definition \ref{Z}, $Z_0 = \mathbb{M}_{2\sigma_z}[\widetilde{X}_0]=0$, which indicates that the transmitter eventually obtains $\sgn{h_1h_2}$.
	
\end{definition}

\begin{definition}
	An $(M, N, P, h_1, h_2, \hat{h}_1, \hat{h}_2, \bigtriangleup_1, \bigtriangleup_2)$-code for the model of Figure \ref{fig2} consists of
	\begin{itemize}
		\item 	Message $W$ is uniformly drawn from  $\mathcal{W} = \lbrace 1,2,..., M\rbrace $.
		
		\item 	At time $i$,  $X_i=g_i(W,\widetilde{\boldsymbol{Y}}^{i-1},\hat{h}_1,\hat{h}_2,\sgn{h_1h_2}$)  ( $i \in\left\lbrace 1, 2, \dots , N\right\rbrace $).
		
		\item At time $i$, the feedback codeword
		$\widetilde{X}_{i}=\widetilde{g}_i(\boldsymbol{Y}^{i},h_1,h_2,\hat{h}_1,\hat{h}_2,
		\bigtriangleup_1,\bigtriangleup_2,\boldsymbol{Z}^{i-1})$.
		
		
		\item At time $N$, the output of the decoder is  $\hat{W}=\psi(\boldsymbol{Y}^{N},h_1,h_2,\hat{h}_1,\hat{h}_2,\bigtriangleup_1,\bigtriangleup_2,\boldsymbol{Z}^{N-1})$. The average decoding error probability is defined the same as that in (\ref{pe}).
		
	\end{itemize}
\end{definition}
A rate $R$ is said to be $(N,\varepsilon,D)$-achievable if for a given coding blocklength  $N$, error probability $\varepsilon$  and a targeted estimation  distortion $D$, there exist encoders and decoders such that
{\small 				\begin{equation}\label{rac}
		\frac{1}{N}\log M\geq {R}-\varepsilon, \,\, P_{e}\leq \varepsilon, \,\, \bigtriangleup_1 \leq D,\,\,\bigtriangleup_2 \leq D.
\end{equation}}%
Similar to Section \ref{model_1}, define the $(N,\varepsilon,D)$-capacity of the TP-QS-FC with I-CSIT and QFC as {\small $\mathcal{C}_\text{I-CSIT}^\text{QFC-TP}(N,\varepsilon,D)$}.


\subsection{ Main results}	
\begin{theorem}\label{th2}
	For given coding blocklength $N$, error probability $\varepsilon$ and targeted estimation distortion $D$,  the $(N,\varepsilon,D)$-capacity {\small $\mathcal{C}_\text{I-CSIT}^\text{QFC-TP}(N,\varepsilon,D)$} of the model of Figure~\ref{fig2} is lower bounded by					
	{\footnotesize 	\begin{align}
			&\mathcal{C}_\text{I-CSIT}^\text{QFC-TP}(N,\varepsilon,D)\geq\mathcal{R}_\text{I-CSIT}^\text{QFC-TP}(N,\varepsilon,D)\nonumber\\
			&\qquad=
			\frac{{N-3}}{2N}\log \left( 1+(H_1+H_2\cdot\sqrt{\rho^{*}})^2\cdot\text{SNR}\cdot\frac{A}{B}\right)\\
			&\quad\quad\quad\quad\quad-								
			\frac{1}{2N}\log \left( \frac{L\cdot\rho_3}{12\cdot ({H_1}^{2}+{H_2}^{2})\cdot\text{SNR}}\right),\nonumber
	\end{align}}
	where	
	{\small 							\begin{eqnarray*}\label{D2}
			H_1&=&[|\hat{h}_{1}|-{D}]^{+},\,\,\,\,\,\,\,\, H_2=[|\hat{h}_{2}|-{D}]^{+},\\
			A&=&(( \sqrt{3\widetilde{P}}-\sigma_z)\cdot[ Q^{-1}( \frac{\varepsilon}{4(N-2)}) ] ^{-1}) ^2,\\
			B&=&(\sqrt{A} +\sigma_z )^2+\frac{3\widetilde{P}\cdot\varepsilon}{2},\,\,\,\,\,\,\, \text{SNR}=\frac{P}{\sigma^{2}},\\
			\rho_3&=&\frac{1}{1+{H_1}^2\cdot\text{SNR}\cdot\frac{A}{B}},\,\,\,\,\, \,L=4[Q^{-1}(\frac{\varepsilon}{4})]^2,
	\end{eqnarray*}}%
	and $\rho^{*}$ is the solution in $[\rho_4, 1)$ of
	\begin{equation*}
		\rho=\frac{1}{1+(H_1+ H_2\cdot\sqrt{\rho})^2\cdot\text{SNR}\cdot\frac{A}{B}}\label{rho2},
	\end{equation*}
	where $\rho_4=\frac{1}{1+(|H_{1}|+|H_{2}|\sqrt{\rho_3})^2\cdot\text{SNR}\cdot\frac{A}{B}}$.
	\begin{IEEEproof}
		See Section \ref{TH2}.
	\end{IEEEproof}	
\end{theorem}	
\begin{remark}\label{sk2-remark-1}
	Note that when $\sigma_{z}$ of the QFC noise and the targeted distortion $D$ tend to zero,
	the model of Figure \ref{fig2} reduces to the TP-QS-FC with P-CSI-TR and NF, and the corresponding rate reduces to
	{\small                                \begin{align}\label{sk2-up}
			\mathcal{R}_\text{P-CSI}^\text{NF-TP}(N,\varepsilon)
			&=
			\frac{{N-3}}{2N}\log \left( 1+(|h_{1}|+|h_{2}|\cdot\sqrt{\rho^{*}})^2\cdot\text{SNR}\right)\nonumber\\
			&\quad\quad\quad-								
			\frac{1}{2N}\log \left( \frac{L\cdot\rho_3}{12\cdot (h_{1}^{2}+h_{2}^{2})\cdot\text{SNR}}\right),		
	\end{align}}%
	where $L=4\left[Q^{-1}\left(\frac{\varepsilon}{2}\right)\right]^2$, $\rho_3=\frac{1}{1+h_1^2\cdot\text{SNR}}$,
	and $\rho^{*}$ is the solution in $[\rho_4, 1)$ of
	{\small \begin{equation}
			\rho\cdot(1+(|h_1|+ |h_2|\cdot\sqrt{\rho})^2\cdot\text{SNR})=1,
	\end{equation}}%
	where $\rho_4=\frac{1}{1+(|h_{1}|+|h_{2}|\sqrt{\rho_3})^2\cdot\text{SNR}}$. Here note that $\mathcal{R}_\text{P-CSI}^\text{NF-TP}(N,\varepsilon)$ can be viewed as a benchmark rate of our proposed scheme.	
\end{remark}
\begin{remark}	\label{sk2-remark-2}										
	Similar to \eqref{H-h} in Remark \ref{R1}, it is not difficult to check that
	{\small 	\begin{equation}\label{H12-h12}
			H_1^{2}\leq h_1^{2},\,\,\,\,\, \,\,\,\,\,  H_2^{2}\leq h_2^{2}.
	\end{equation}	}%
\end{remark}

	%

\subsection{Numerical  results}		
\begin{figure}[htb]
	\centering
	\includegraphics[scale=0.2]{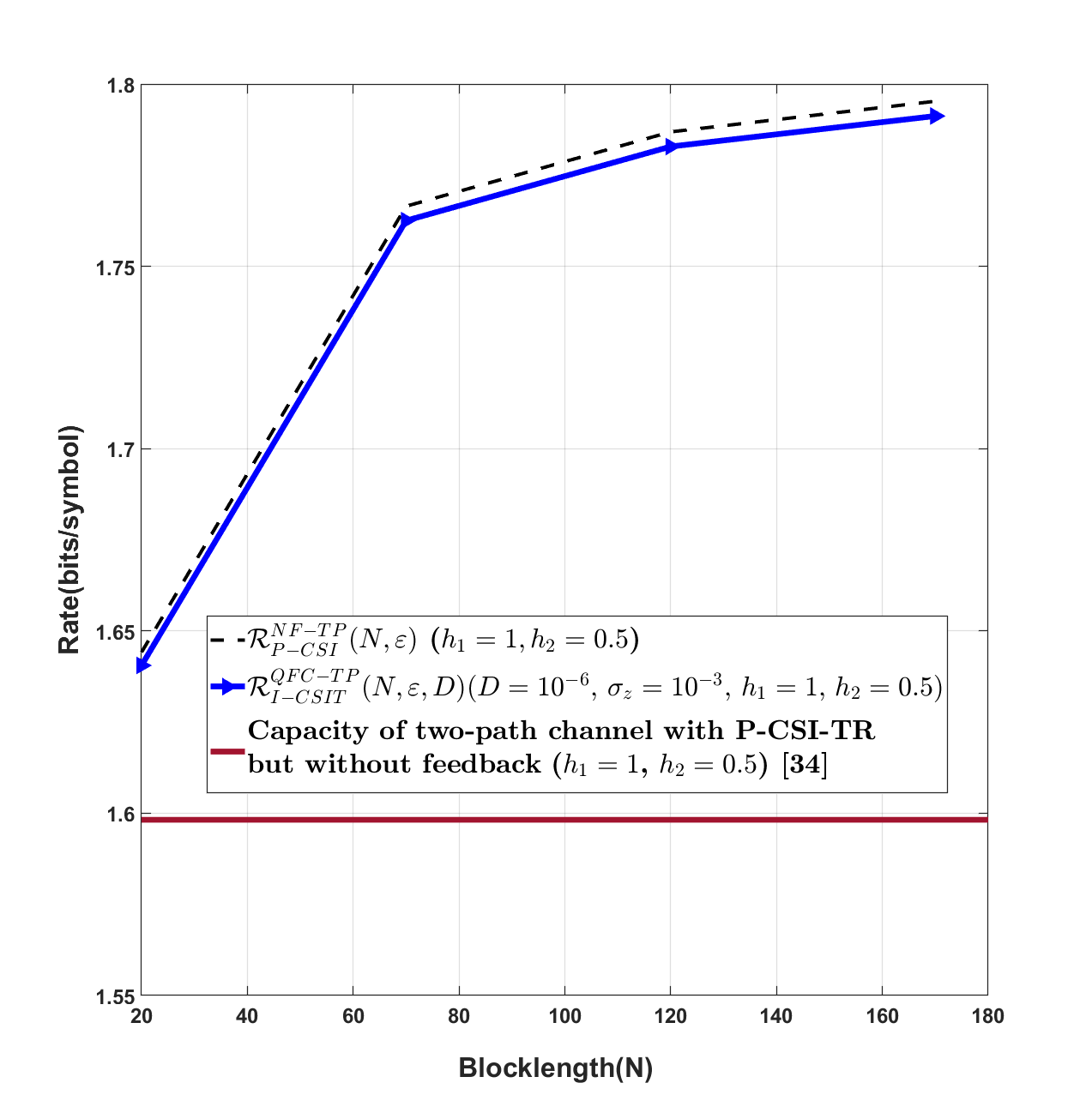}
	\caption{ { Achievable rate versus coding blocklength $N$ and channel coefficients $h_1$, $h_2$ under $\text{SNR}=10$, $\widetilde{P}=10$, $\varepsilon=10^{-6}$, $D=10^{-6}$, and $\sigma_z=10^{-3}$}}
	\label{fg5}
\end{figure}
As shown in Figure~\ref{fg5}, when $D$ and $\sigma_z$ are small, the rate of our scheme almost approaches the benchmark rate of (\ref{sk2-up}). Furthermore,
Figure~\ref{fg5} also demonstrates that feedback may enhance the transmission rate of the TP-QS-FC with P-CSI-TR.

\vspace{-10pt}
\subsection{Proof of Theorem \ref{th2}}\label{TH2}

In the proof of Theorem \ref{th1}, an SK-type scheme has been proposed for the fading channel with I-CSIT and QFC. To build up a similar SK-type scheme for the two-path case, we
treat the signal of the second path as a relay, and combine the amplify-and-forward relay strategy with the previously proposed scheme of Theorem \ref{th1}. The detail of our scheme is explained below.


%
%
%

Analogous to the previous sections, the message $W$ is mapped to the midpoint $\theta$ of $W$'s sub-interval.

$\textbf{Initialization:}$ At time instant $1$, the transmitter encodes $\theta$ as $X_{1}=\sqrt{{12P}}\cdot \theta$, resulting in $Y_{1}=h_1\cdot X_{1}+\eta_{1}$.

At time instant $2$, the transmitter sends $X_{2}=0$,
leading to $Y_{2}=h_2\cdot X_{1}+h_1\cdot 0+\eta_{2}=h_2\cdot X_{1}+\eta_{2}$.

Then the receiver computes his first estimation $\hat{\theta}_{1}$ of $\theta$ by
{\small 	\begin{align}
		&\hat{\theta}_{1}=\kappa\cdot\frac{Y_{1}}{h_1\cdot\sqrt{12P}}+(1-\kappa)\cdot\frac{Y_{2}}{h_2\cdot\sqrt{12P}}\nonumber\\	
		&=\theta+\kappa\cdot\frac{\eta_{1}}{h_1\cdot\sqrt{12P}}+(1-\kappa)\cdot\frac{\eta_{2}}{h_2\cdot\sqrt{12P}}
		=\theta+\epsilon_1,
\end{align}}%
where $\epsilon_1=\hat{\theta}_{1}-\theta$, and the weighting coefficient $\kappa\in [0,1]$ is chosen to minimize $\mathbb{E}[(\epsilon_1)^{2}]$.

Similar to MMSE, it is not difficult to check that the minimum $E[(\epsilon_1)^{2}]=\frac{\sigma^2}{12P\cdot(h_1^2+h_2^2)}$, which is achieved by $\kappa = \frac{h_1^{2}}{h_1^{2}+h_2^{2}}$.
The  receiver sets $\hat{\theta}_{2}=\hat{\theta}_{1}$, so that  $\epsilon_2=\hat{\theta}_{2}-\theta=\epsilon_1$, and then he sends
{\small 		\begin{equation}
		\widetilde{X}_2=\mathbb{M}_{\widetilde{d}}[\gamma_{2}\cdot\hat{\theta}_{2}+V_2]
\end{equation}}%
back to the transmitter via the QFC, where $\widetilde{d}=\sqrt{12\widetilde{P}}$, $V_{2}$ is a dither signal uniformly distributed in $[-\frac{\widetilde{d}}{2},\frac{\widetilde{d}}{2})$, it is independent of all other random variables and known by all parties, and $\gamma_{2}$ will be defined later. The feedback signal received by the transmitter is denote by $\widetilde{Y}_{2}=\widetilde{X}_2+Z_{2}$.

At time instant $3$, the transmitter sends
{\small \begin{align}\label{X2_11}
		X_{3}&=\sgn{h_1h_2}\cdot\alpha\cdot\mathbb{M}_{\widetilde{d}}[\widetilde{Y}_{2}-\gamma_{2}\cdot\theta-V_2]\nonumber\\
		&\mathop{=}\limits^{\text{(a)}}\sgn{h_1h_2}\cdot\alpha\cdot\mathbb{M}_{\widetilde{d}}[\gamma_{2}\cdot\epsilon_{2}+Z_2],
\end{align}}%
where $\text{(a)}$ follows from Property (ii) of Proposition \ref{M_d-p},  and $\alpha$ will be defined later.
Once receiving
{\small \begin{equation}\label{sk2-y3}
		Y_{3}=h_{1}\cdot X_{3}+h_{2}\cdot 0+\eta_{3}=h_{1}\cdot X_{3}+\eta_{3},
\end{equation}}%
the receiver calculates an auxiliary signal
{\small \begin{equation}\label{sk2-aux1}
		\dot{Y}_{3}=Y_{3}-\sgn{h_1h_2}\cdot h_{1}\cdot\alpha\cdot Z_2.
\end{equation}}%
and updates his estimation  $\hat{\theta}_{2}$ of $\theta$ by using
this auxiliary signal, i.e.,
{\small 	\begin{equation}\label{sk2-beta1}
		\hat{\theta}_{3}=\hat{\theta}_{2}-\beta_{2}\cdot\dot{Y}_{3},
\end{equation}}%
where $\beta_{2}$ will be defined later.	Then the receiver sends
{\small \begin{equation}\label{sk2-rx3}
		\widetilde{X}_3=\mathbb{M}_{\widetilde{d}}[\gamma_{3}\cdot\hat{\theta}_{3}+V_3]
\end{equation}}%
back to the transmitter, where $V_3$ is a dither signal defined the same as $V_2$, and $\gamma_{3}$ will be defined later.
The feedback signal received by the transmitter is denote by $\widetilde{Y}_{3}=\widetilde{X}_3+Z_{3}$.

$\textbf{Iteration:}$
At time instant $k \in \left\lbrace 4,\ldots,N\right\rbrace$, after receiving the feedback $\widetilde{Y}_{k-1}$, the transmitter sends
{\small 							\begin{align}\label{sk2-alpha}
		X_{k}&=[\sgn{h_1h_2}]^{k-2}\cdot\alpha\cdot \mathbb{M}_{\widetilde{d}}[\widetilde{Y}_{k-1}-\gamma_{k-1}\cdot\theta-V_{k-1}]\nonumber\\
		&=[\sgn{h_1h_2}]^{k-2}\cdot\alpha\cdot\mathbb{M}_{\widetilde{d}}[\gamma_{k-1}\cdot \epsilon_{k-1}+Z_{k-1}],
\end{align}}%
where $V_{k-1}$
is a dither signal defined the same as those stated above, and $\gamma_{k-1}$ will
be defined later.

Once receiving
{\small 	\begin{equation}\label{sk2-yi}
		Y_{k}=h_{1}\cdot X_{k}+h_{2}\cdot X_{k-1}+\eta_{k},
\end{equation}}%
the receiver calculates the auxiliary signal
{\small 	\begin{align}\label{sk2-aux2}
		\dot{Y}_{k}=&Y_{k}-[\sgn{h_1h_2}]^{k-2}\cdot h_{1}\cdot\alpha\cdot Z_{k-1}\nonumber\\
		&-[\sgn{h_1h_2}]^{k-3}\cdot h_{2}\cdot\alpha\cdot ( Z_{k-2}+\beta_{k-2}\cdot\dot{Y}_{k-1}),
\end{align}}%
and updates his estimation  $\hat{\theta}_{k}$ of $\theta$ by
{\small \begin{equation}\label{sk2-beta2}
		\hat{\theta}_{k}=\hat{\theta}_{k-1}-\beta_{k-1}\cdot\dot{Y}_{k}.
\end{equation}}%
Define $\epsilon_{n}=\hat{\theta}_{n}-\theta$ ($n\in\{1,\cdots,N\}$), then (\ref{sk2-beta2}) yields that
{\small \begin{equation}\label{sk2-ep-r}
		\epsilon_{k}=\epsilon_{k-1}-\beta_{k-1}\cdot\dot{Y}_{k}.
\end{equation}}%
The receiver sends $\widetilde{X}_{k}$ back to the transmitter via the QFC, where
{\small \begin{equation}\label{sk2-rxk}
		\widetilde{X}_{k}=\mathbb{M}_{\widetilde{d}}[\gamma_{k}\cdot\hat{\theta}_{k}+V_{k}].
\end{equation}}%

$\textbf{Decoding:}$
The receiver's final decoding
process is the same as that of Section \ref{sk0}, and we omit it here.					
\subsubsection{ Performance analysis} \hfil	

First, we construct a coupled system which is in analogous to that of Theorem~\ref{th1}.
Then, for {\small  $i \in \left\lbrace {2},\ldots,N-1\right\rbrace$}, we define {\small $E_i$} as a modulo-aliasing error event, and it is given by
{\small \begin{eqnarray}\label{Ei2}	
		E_i\triangleq\lbrace \gamma_{i}\cdot\epsilon_{i}+Z_{i}\notin [ -\frac{\widetilde{d}}{2},\frac{\widetilde{d}}{2}) \rbrace.
\end{eqnarray}}%
Furthermore, we define $E_{N}$ as the receiver's final decoding error event, which is given by
{\small \begin{eqnarray}\label{sk2-EN}	
		E_{N}\triangleq\lbrace \epsilon_{N} \notin [ -\frac{1}{2^{NR+1}},\frac{1}{2^{NR+1}}) \rbrace.
\end{eqnarray}}%
In parallel,
let {\small  $E_i^{'} \triangleq \left\lbrace \gamma_{i}\cdot\epsilon'_{i}+Z_{i} \notin \left[-\frac{\widetilde{d}}{2},\frac{\widetilde{d}}{2}\right)\right\rbrace$} and {\small  $E_{N}^{'}\triangleq\left\lbrace \epsilon'_{N} \notin \left[ -\frac{1}{2^{NR+1}},\frac{1}{2^{NR+1}}\right) \right\rbrace$} be the corresponding error events of the coupled system.	

Now we establish an upper bound on the decoding error probability $\Pe$ of the original system, see the following Lemma \ref{sk2-lemma1}.

\begin{lemma}\label{sk2-lemma1}
	The  decoding error probability $\Pe$ in the original system can be upper bounded by
	{\footnotesize 	\begin{equation}\label{sk2-lemma1-1}
			P_e \! \leq\!\sum_{i=2}^{N-1}\! 2Q(\!\frac{\sqrt{3\widetilde{P}}\! -\! \sigma_z}{\sqrt{\gamma_i^2  \alpha_{i}'(H_1,H_2)}} \!)
			\! +\!2\! \mathop{Q}(\! \frac{1}{2^{NR+1}\sqrt{\alpha_{N}'(H_1,H_2)}}\!),
	\end{equation}}%
	where
{\footnotesize  \begin{align}\label{sk2-lemma1-2}
	&\alpha_{2}'(H_{1},H_{2})=\frac{\sigma^2}{ 12P\cdot(H_1^2+H_2^2)},\nonumber\\
	&\alpha_{3}'(H_{1},H_{2})=\alpha_{2}'(H_{1},H_{2})
	\cdot({1+\frac{H_1^2\cdot\alpha^2\cdot\gamma_2^2
			(H_{1},H_{2})}{\sigma^2}})^{-1}, \\		
	&\alpha_{j}'(H_{1},H_{2})=\alpha_{j-1}'(H_{1},H_{2})\nonumber\\
	&\quad\times({1+\frac{\alpha^2\cdot(|H_{1}|\gamma_{j-1}+|H_{2}|\gamma_{j-2})^2\cdot \alpha_{j-1}'(H_{1},H_{2})}{\sigma^2}})^{-1},\nonumber	
	\end{align}}
	and $j\in\{4,5,...,N\}$.
\end{lemma}
\begin{IEEEproof}
See \textbf{Appendix}~\ref{b-0}.
\end{IEEEproof}

Note that the above upper bound on $\Pe$ depends on $\gamma_i$ and $\alpha$, which are determined as follows.

\emph{Determination of {\small $\gamma_i$} and {\small $\alpha$}}: Analogous to that of the proof of Theorem \ref{th1},
to ensure {\small $P_e$ } of (\ref{sk2-lemma1-1}) satisfying {\small $P_e \leq \varepsilon$}, let
{\small \begin{equation}\label{sk2-pe4}
2\mathop{Q}( \frac{1}{2^{NR+1}\sqrt{\alpha_{N}'(H_1,H_2)}})=\frac{\varepsilon}{2},
\end{equation}}%
and
{\small \begin{equation}\label{sk2-pe5}
2Q(\frac{\sqrt{3\widetilde{P}} - \sigma_z}{\sqrt{\gamma_i^2 \cdot \alpha_{i}'(H_1,H_2)}})	=p'_m=\frac{\varepsilon}{2(N-2)},
\end{equation}}%
for $i\in\{2,\cdots,N-1\}$, which indicates that
{\small \begin{equation}\label{sk2-pe6}
\sum_{i=2}^{N-1}2Q(\frac{\sqrt{3\widetilde{P}} - \sigma_z}{\sqrt{\gamma_i^2 \cdot \alpha_{i}'(H_1,H_2)}} )=\frac{\varepsilon}{2}.
\end{equation}}%
Then from (\ref{sk2-pe5}), we have
{\small \begin{equation}\label{sk2-pe7}
\gamma_i = \sqrt{\frac{A}{\alpha_{i}'(H_1,H_2)}},
\end{equation}}%
where {\small $A=(( \sqrt{3\widetilde{P}}-\sigma_z)\cdot[Q^{-1}(\frac{p'_m}{2})]^{-1}) ^2$}.

On the other hand, note that $\alpha$ is a parameter used to guarantee the transmitting power of the original system, then from (\ref{sk2-pe7}),
it is not difficult to show that if we set
{\small \begin{equation}\label{sk2-alpha1}
\alpha=\sqrt{\frac{P}{B}},
\end{equation}}%
where {\small $B=(\sqrt{A} +\sigma_z )^2+\frac{3\widetilde{P}\cdot\varepsilon}{2}$}, the power constraint
{\small $\mathbb{E}[(X_{i+1})^2]\leq P$}
is ensured.

To this end, it remains to determine $\gamma_i$, which depends on $\alpha_{i}'(H_1,H_2)$.
Interestingly, we find that substituting (\ref{sk2-alpha1}) and (\ref{sk2-pe7})
into (\ref{sk2-lemma1-2}), $\alpha_{i}'(H_1,H_2)$ is further given by
{\footnotesize  \begin{align}
\alpha_{2}'(H_{1},H_{2})&=\frac{\sigma^2}{ 12P\cdot(H_1^2+H_2^2)},\label{sk2-conver-1}\\
\alpha_{3}'(H_{1},H_{2})&=\frac{\sigma^2}{ 12P\cdot(H_1^2+H_2^2)}\cdot\frac{1}{1+{H_1^2\cdot\frac{P}{\sigma^{2}}\cdot\frac{A}{B}}},\label{sk2-conver-2}  \\
\alpha_{i}'(H_{1},H_{2})&=\frac{\alpha_{i-1}'(H_{1},H_{2})}{1+(|H_{1}|+|H_{2}|\sqrt{\frac{\alpha_{i-1}'(H_{1},H_{2})}
		{\alpha_{i-2}'(H_{1},H_{2})}})^2\cdot\frac{P}{\sigma^{2}}\cdot\frac{A}{B}},\label{sk2-conver}
\end{align}}%
where  $i \in \{4,\cdots,N\}$.
To further calculate the general term of $\alpha_{i}'(H_{1},H_{2})$,
define {\small $\rho_{i}\triangleq\frac{\alpha_{i}'(H_{1},H_{2})}{\alpha_{i-1}'(H_{1},H_{2})}$},
then (\ref{sk2-conver}) can be re-written as
{\small \begin{equation}\label{sk2_rho}
\rho_i=\frac{1}{1+(|H_{1}|+|H_{2}|\sqrt{\rho_{i-1}})^2\cdot \frac{P}{\sigma^{2}}\cdot\frac{A}{B}}.
\end{equation}}%
The following Lemma \ref{sk2-lemma-3} proves the convergence of {$\rho_i$}.

\begin{lemma}\label{sk2-lemma-3}
The series $\{\rho_i\}$ converges to $\rho^{*}$, where $\rho^{*}$ is the solution in $[\rho_4,1)$ of
{\small 	\begin{equation}\label{sk2-l3}
	\rho=\frac{1}{1+(|H_{1}|+|H_{2}|\sqrt{\rho})^2\cdot\frac{P}{\sigma^{2}}\cdot\frac{A}{B}},
	\end{equation}}%
\end{lemma}
\begin{IEEEproof}	
The proof of Lemma \ref{sk2-lemma-3} includes:
1) {$\rho_i$} converges to $\rho^{*}$, and 2)
for $[\rho_4,1)$, there exists a $\rho^{*}$ satisfying (\ref{sk2-l3}).

Here note that if such a $\rho^{*}$ exists, choose $\rho_4=\rho^{*}$, then it is easy to check that
$\rho_i=\rho^{*}$ for $i \in \{4,\cdots,N\}$, which indicates that {$\rho_i$} converges to $\rho^{*}$. Now it remains to prove that such a $\rho^{*}$ does exist, see the detail below.

First, it is easy to check that when $\rho = 1$, $1-\frac{1}{1+(|H_{1}|+|H_{2}|)^2\cdot\frac{P}{\sigma^{2}}\cdot\frac{A}{B}}\geq 0$.
Then observing that
{\footnotesize  \begin{align}
	\rho_3 &=\frac{\alpha_{3}'(H_{1},H_{2})}{\alpha_{2}'(H_{1},H_{2})}= \frac{1}{1 + H_1^2\cdot\frac{P}{\sigma^{2}}\cdot\frac{A}{B}},\label{sk2-r3r4}\\
	\rho_4 &=\frac{\alpha_{4}'(H_{1},H_{2})}{\alpha_{3}'(H_{1},H_{2})} =\frac{1}{1 + (|H_1| + |H_2|\sqrt{\rho_3})^2\cdot\frac{P}{\sigma^{2}}\cdot\frac{A}{B}},\label{sk2-r3r5}
	\end{align}}
	and we can easily check that $\rho_3 \geq \rho_4$, which yields that
	{\small \begin{align}\label{rho1-vs-rho2_2}
	\rho_4&=\frac{1}{1 + (|H_1| + |H_2|\sqrt{\rho_3})^2 \cdot \frac{P}{\sigma^{2}} \cdot \frac{A}{B}} \nonumber\\
	&\leq \frac{1}{1 + (|H_1| + |H_2|\sqrt{\rho_4})^2 \cdot \frac{P}{\sigma^{2}} \cdot \frac{A}{B}},
	\end{align}}%
	namely, {\small $\rho_4-\frac{1}{1 + (|H_1| + |H_2|\sqrt{\rho_4})^2 \cdot \frac{P}{\sigma^{2}} \cdot \frac{A}{B}}\leq 0$}, which indicates that such a {\small $\rho^{*}\in[\rho_4,1)$} does exist, and the proof is completed.
\end{IEEEproof}

Lemma \ref{sk2-lemma-3} shows that $\{\rho_i\}$ converges to $\rho^{*}$, however, we still do not know the specific values of $\{\rho_i\}$ before approaching its steady point, which indicates that
(\ref{sk2-conver}) cannot be determined for $\{\rho_i\}$ not approaching its steady point $\rho^{*}$. In order to solve this problem, an interesting trick proposed in \cite{ozarow} is
used to guarantee $\rho_{i}=\rho^{*}$ for  $i\in \{4,\cdots,N\}$, which is described below.

In the encoding-decoding procedure stated above, at time $4$ (beginning of the iteration), an artificial noise $U\sim\mathcal{N}(0,\sigma^{2}_0)$, which is known by all parties and is independent of all random variables introduced above, is added to both the encoding function (\ref{sk2-alpha}) and the auxiliary signal in (\ref{sk2-aux2}) which is the key to the decoding function. To this end, in such a case in the presence of artificial noise, $\alpha_{4}^{u}(H_{1},H_{2})$ (analogously defined as (\ref{sk2-conver})) can be calculated by
{\small \begin{equation}\label{alpha_4}
\alpha_{4}^{u}(H_{1},H_{2})=\frac{\alpha'_{3}(H_{1},H_{2})}{1+(|H_{1}|+|H_{2}|\sqrt{\rho_{3}})^2
	\cdot\frac{P}{\sigma^2+\sigma_0^{2}}\cdot\frac{A}{B}},
\end{equation}}%
and the corresponding $\rho_{4}^{u}\triangleq\frac{\alpha_{4}^{u}(H_{1},H_{2})}{\alpha'_{3}(H_{1},H_{2})}$ (here $U$ is added at time $4$, hence the previous $\alpha'_{3}(H_{1},H_{2})$ remains the same) is calculated by
{\small \begin{equation}\label{sk2-r4}
\rho_4^{u}=\frac{1}{1+(|H_{1}|+|H_{2}|\sqrt{\rho_{3}})^2
	\cdot\frac{P}{\sigma^2+\sigma_0^{2}}\cdot\frac{A}{B}}.
\end{equation}}%
From (\ref{sk2-r4}), we see that $\rho_{4}^{u}$ depends on $\sigma^{2}_0$ (variance of the artificial noise $U$), and it can be easily check that when $\sigma^{2}_0=0$, $\rho_{4}^{u}=\rho_{4}$ (see (\ref{sk2-r3r5})), and when $\sigma^{2}_0\rightarrow \infty$,
$\rho_{4}^{u}\rightarrow 1$, which implies that $\rho_{4}^{u}\in [\rho_{4}, 1)$.
Since the steady point $\rho^{*}$ of $\{\rho_i\}$ also takes values in $[\rho_{4}, 1)$, by choosing an appropriate $\sigma^{2}_0$, we can set $\rho_{4}^{u}=\rho^{*}$. Furthermore, since $U$ is not added in next time instants, it is not difficult to check that (\ref{sk2_rho}) still holds for $i\in\{5,6,...,N\}$, namely,
{\small \begin{equation}\label{sk2_rho-1}	
\rho_i^{u}=\frac{1}{1+(|H_{1}|+|H_{2}|\sqrt{\rho_{i-1}^{u}})^2\cdot\frac{P}{\sigma^{2}}\cdot\frac{A}{B}}.
\end{equation}}%
Observing that
{\small \begin{equation}\label{sk2_rho-2}
\rho^{*}=\frac{1}{1+(|H_{1}|+|H_{2}|\sqrt{\rho^{*}})^2\cdot\frac{P}{\sigma^{2}}\cdot\frac{A}{B}},
\end{equation}}%
and $\rho_{4}^{u}=\rho^{*}$, we conclude that $\rho_i^{u}=\rho_{4}^{u}=\rho^{*}$ for $i\in\{5,6,...,N\}$, and this implies that
{\footnotesize \begin{align}\label{sk2-conver2-i}
\alpha_{i}'(H_{1},H_{2})=&\frac{\rho_3}{ 12\cdot\frac{P}{\sigma^{2}}\cdot(H_1^2+H_2^2)}\nonumber\\
&\times( \frac{1}{1+(|H_{1}|+|H_{2}|\sqrt{\rho^{*}})^2\cdot\frac{P}{\sigma^{2}}\cdot\frac{A}{B}}) ^{i-3},
\end{align}}%
where $\rho_3$ is defined in (\ref{sk2-r3r4}).

Now substituting (\ref{sk2-conver-1}) (\ref{sk2-conver-2}) and (\ref{sk2-conver2-i}) into (\ref{sk2-pe7}),
$\gamma_i$ is determined.

Finally, substituting (\ref{sk2-conver-1}) (\ref{sk2-conver-2}) and (\ref{sk2-conver2-i}) into (\ref{sk2-pe4}),
the rate $R$ can be re-written by
{\footnotesize  \begin{align}\label{sk2-r}
R&=
\frac{{N-3}}{2N}\log \left( 1+(|H_1|+|H_2|\sqrt{\rho^{*}})^2\cdot\text{SNR}\cdot\frac{A}{B}\right)\\
&\quad\quad\quad\quad-								
\frac{1}{2N}\log ( \frac{L\cdot\rho_3}{12\cdot ({H_1}^{2}+{H_2}^{2})\cdot\text{SNR}}),\nonumber
\end{align}}
where {\small $L=4\left[Q^{-1}\left(\frac{\varepsilon}{4}\right)\right]^2$}.

In addition, substituting (\ref{sk2-ep1}) and (\ref{sk2-dotY}) into (\ref{sk2-beta}), and applying
(\ref{sk2-alpha1}) and (\ref{sk2-pe7}) to (\ref{sk2-beta}), the MMSE coefficient $\beta_{i}$
can be determined. The proof of Theorem \ref{th2} is completed.

\section{{The multi-Path Quasi-Static Fading Channel with Noiseless Feedback}}\label{sk3}	

{Recall that in the two-path quasi-static fading channel model with I-CSIT and QFC, the achievable rate is given by introducing a co-efficient factor $\rho^{*}$ which characterizes the correlation between the codewords of the transmitter and the relay, and $\rho^{*}$ is the unique solution of an equation. However, this relay based scheme cannot be extended to arbitrary multi-path fading case since the co-efficient factor $\rho^{*}$ determined by an equation will be expanded to multi factors (for $m$-path fading, there are approximately $\frac{m(m-1)}{2}$ factors, where $m\geq 2$), and these factors
are determined by different equations related with all factors, which may not hold since we cannot guarantee all these
equations have solutions.
To this end, in this section, we propose an SK-type scheme for arbitrary multi-path fading channel with noiseless feedback, which provides a different way to design SK-type scheme for practical wireless communication scenarios.}

\subsection{{System model and main result}}\label{model_3}
\begin{figure}[htb]
\centering
\includegraphics[scale=0.18]{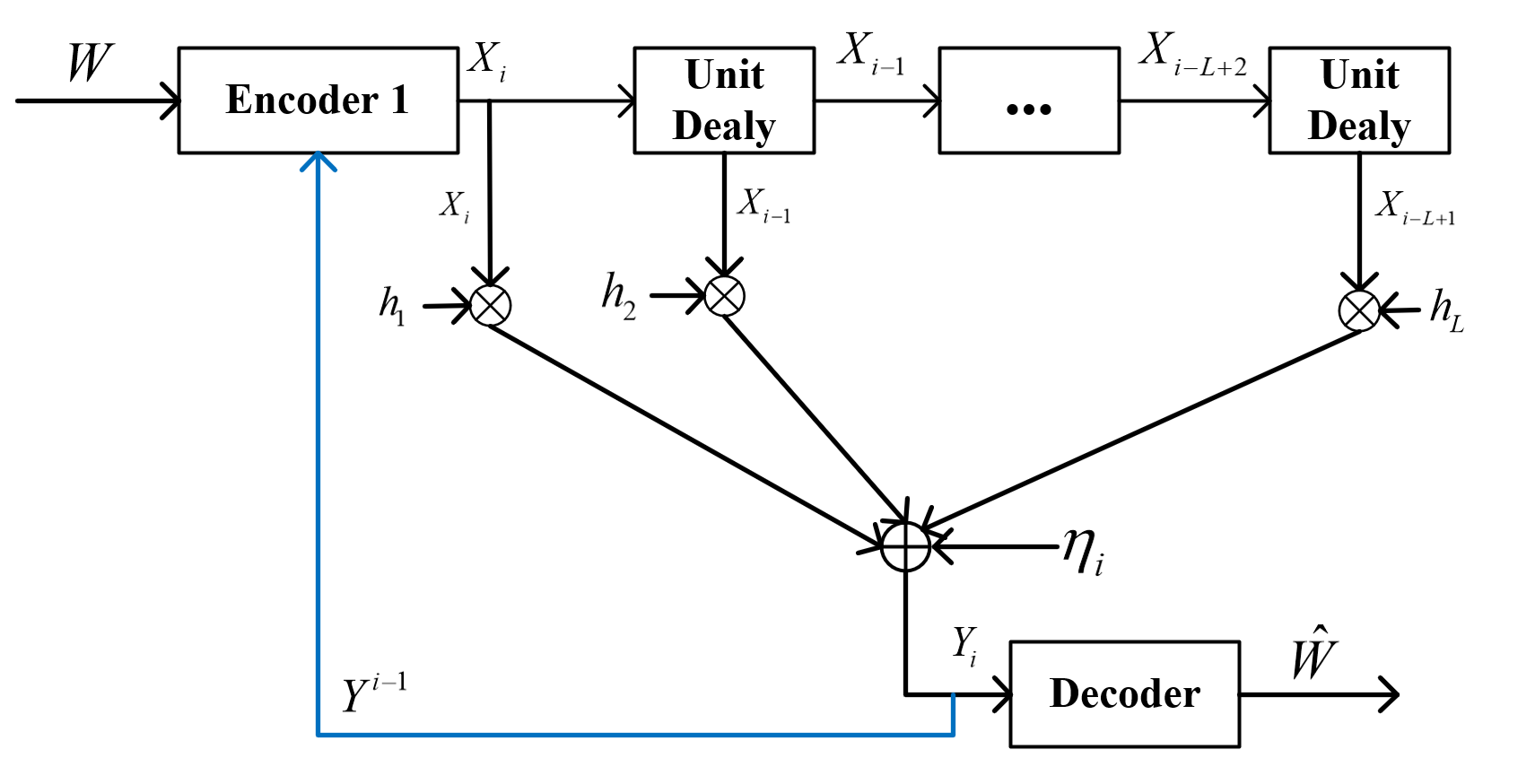}
\caption{{L-path channel with noiseless feedback}}
\label{path-1}
\end{figure}

{In Figure \ref{path-1}, at time $i\in\left\lbrace 1,2,\dots,N\right\rbrace $, the channel input-output is given by
\begin{equation}\label{model3-1}
Y_i=\sum_{l=1}^{L} h_l\cdot X_{i-l+1}+\eta_{i},	
\end{equation}
where $L$ ($L\geq 2$) is the number of fading paths, the channel input $X_i\in \mathbb{C}$ satisfies average power constraint $\frac{1}{N}\sum\nolimits_{i=1}^{N}\mathbb{E}[\|X_i\|^{2}]\leq P$,  $\mathbf{h}\triangleq[h_1,\dots, h_L]^{T}\in \mathbb{C}^{L}$
is the quasi-static fading coefficient, namely, $\mathbf{h}$  remains a constant vector within a block, and it is perfectly known by the transceiver, $\eta_i\sim\mathcal{CN}(0,\sigma^{2}) $ is the complex Gaussian noise which is independent and identically distributed (i.i.d.) across the time index $i$, and $Y_i$ is the output of the forward channel.}

{	
\begin{definition} An $(|\mathcal{W}|, N, P, \mathbf{h})$-code for the model of Figure \ref{path-1} consists:
\begin{itemize}
	\item 	The message $W$ is uniformly drawn from $\mathcal{W} = \left\lbrace 1, 2, \dots,  2^{NR}\right\rbrace $.
	\item 	At time $i$ ($i \in\left\lbrace 1, 2, \dots , N\right\rbrace $),
	the codeword $X_{i}=g_i(W,\boldsymbol{Y}^{i-1}, \mathbf{h})$, where
	$g_i$ is the encoding function, and $\boldsymbol{Y}^{i-1}=[Y_1,\dots,Y_{i-1}]^{T}$ is the feedback signal at previous time instants.		
	\item At time $N$, the output of the decoder is  $\hat{W}=\psi(\boldsymbol{Y}^{N},\mathbf{h})$. The average decoding error probability is defined the same as that in (\ref{pe}).
\end{itemize}
\end{definition}
The rate ${R}$ is said to be $(N,\varepsilon)$-achievable if for given coding blocklength  $N$, and error probability $\varepsilon$, there exist encoders and decoders such that
\begin{equation}
\frac{1}{N}\log |\mathcal{W}| \geq {{R}}-\varepsilon, \,\,\,\,\,\, P_{e}\leq \varepsilon.
\end{equation}
The $(N,\varepsilon)$-capacity is  the
supremum over all $(N,\varepsilon)$-achievable rates defined above, and
it is denoted by  $\mathcal{C}_\text{L-Path}(N,\varepsilon)$, e.g., when $L=3$, it is denoted by
$\mathcal{C}_\text{3-Path}(N,\varepsilon)$, and its lower bound is denoted by $\mathcal{R}_\text{3-Path}(N,\varepsilon)$.}
{
\subsection{ Main result}		
\begin{theorem}\label{th3}	
For given coding blocklength $N$  and error probability $\varepsilon$,  the $(N,\varepsilon)$-capacity $\mathcal{C}_\text{L-Path}(N,\varepsilon)$ is lower bounded by
\begin{align}
	&\mathcal{C}_\text{L-Path}(N,\varepsilon)\geq\mathcal{R}_\text{L-Path}(N,\varepsilon)\nonumber\\
	&=\max\limits_{K\in \{L,\cdots,N-L+1\}}(\sum_{k=1}^{K}(
	\frac{{\Phi-1}}{N}\log (1+\|H_k\|^2\cdot\frac{P_k}{\sigma^{2}})	\nonumber	\\
	&\quad\quad\quad\quad\quad\quad\quad\quad\quad\quad-\frac{1}{N}\log (\frac{\Xi}{12\cdot \|H_k\|^{2}\cdot\frac{P_k}{\sigma^{2}}}))),				
\end{align}
where $\mathbf{H} \triangleq [H_1, \cdots, H_K]^{T}$ is the Discrete Fourier Transform (DFT)  of $\mathbf{{\tilde{h}}}=[h_1, \cdots, h_L, \underset{{K-L \,\,\text{zeros}}}{\underbrace{0,\dots,0}} ]^{T}$, i.e.,
\[
{H}_{k} = \sum_{n=1}^{K}  e^{-2\pi j \cdot\frac{(n-1) (k-1)}{K}} \cdot\mathbf{{\tilde{h}}}_{n},\,\,\,\,\,\,  1 \leq k\leq K,
\]
$\Xi=4\left[Q^{-1}\left(\frac{\varepsilon}{4K}\right)\right]^2$, $\Phi= \lfloor\frac{N}{L+K-1}\rfloor$,
$P_1,\cdots,P_K$ are the power coefficients solved by the water-filling algorithm			
\begin{equation}
	P_k=\max(q-\frac{\sigma^2}{\|H_k\|^{2}},0),  \,\,\,\, 1 \leq k\leq K
\end{equation}
and $q$ is chosen so that
\begin{equation}
	\sum_{k=1}^{K}\max(q-\frac{\sigma^2}{\|H_k\|^{2}},0)=KP.
\end{equation}
\end{theorem}}	
\subsection{Numerical  results}	
\begin{figure}[htb]
\centering
\includegraphics[scale=0.20]{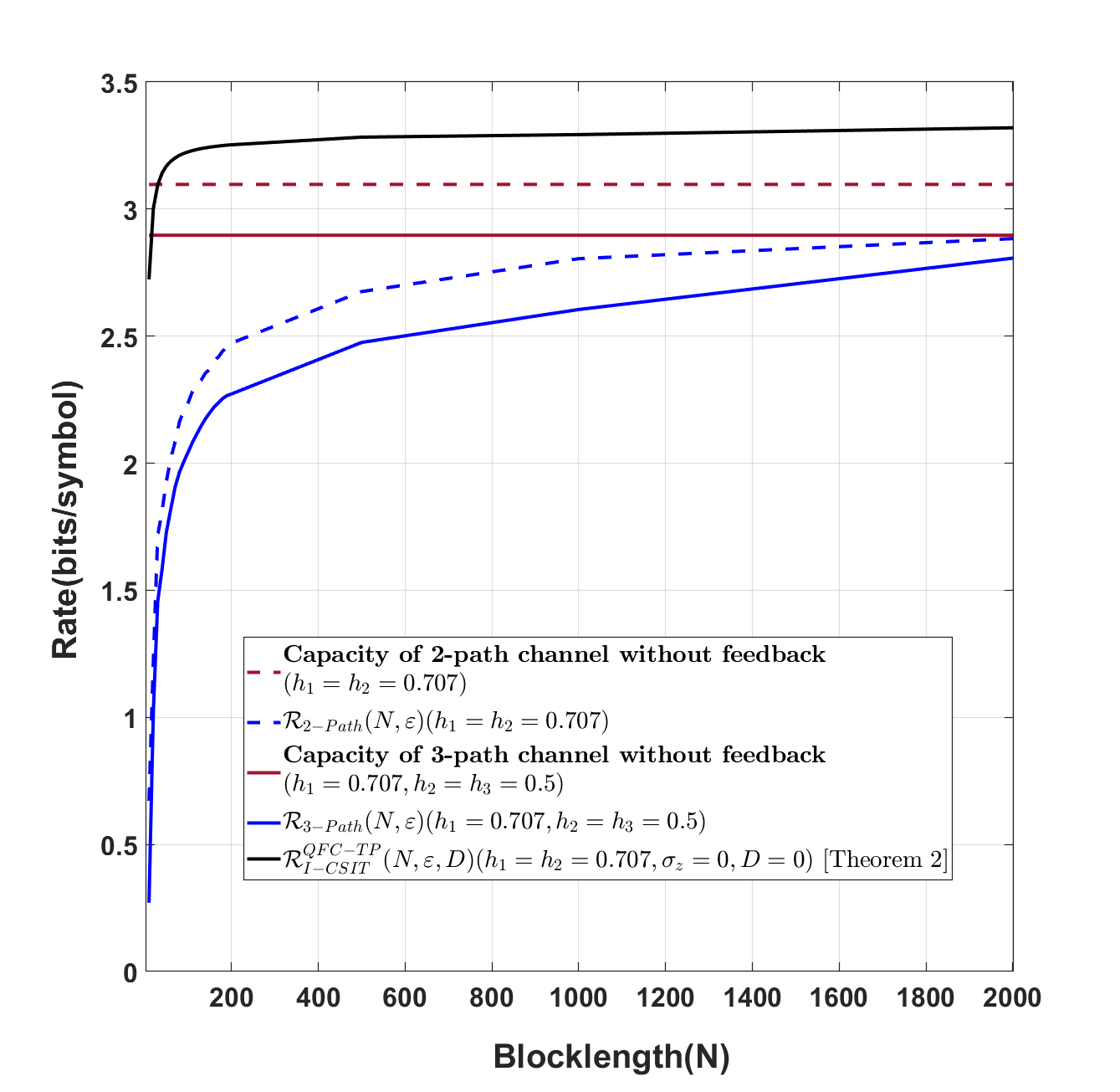}
\caption{{ Rate versus coding blocklength $N$ for $\text{SNR} =10$ and $\varepsilon=10^{-4}$}}
\label{path-2}
\end{figure}
{	
Figure \ref{path-2} shows that our new feedback scheme for the multi-path fading channel performs
worse than the relay based SK-type scheme of Theorem \ref{th2}, and this is because in our new scheme, we add additional redundancy into the codeword which helps to improve the coding performance in the presence of multi-path fading. Furthermore, our scheme cannot even asymptotically approach the capacity of the same model without feedback \cite{memory}, which indicates that our scheme is not ``good'' in general.}

{
\subsection{Proof of Theorem \ref{th3}}\label{TH3}	
The message is split into $K$ sub-messages ($K\in \{L,\cdots,N-L+1\}$), and each sub-message is transmitted over a certain block
with blocklength $L+K-1$, where $L$ is the number of fading paths. Hence
the total codeword length $N$ is divided into $\Phi\triangleq \lfloor\frac{N}{L+K-1}\rfloor$ blocks.
To effectively decrease the impact of inter-symbol interference (ISI) induced by multi-path delay spread on the coding performance, each block is designed with a cyclic convolution property \cite{ofdm}. This enables each frequency-domain signal to be treated as being transmitted over independent parallel sub-channels, and then we adopt the SK-type scheme to each sub-channel.}

{
\emph{Message splitting and mapping}:											
The original message \( W \in \mathcal{W} \), with \( |\mathcal{W}| = 2^{NR} \), is split into \( K \) independent sub-messages \( W = (W_1, W_2, \ldots, W_K) \). Each sub-message \( W_k \in \mathcal{W}_k \) satisfies \( |\mathcal{W}_k| = 2^{NR_k} \) for \( k \in\{ 1, 2, \ldots, K \}\), and the total message space satisfies
\[
\prod_{k=1}^{K} |\mathcal{W}_k| = |\mathcal{W}|,
\]
also indicates that
\begin{equation}\label{w2}
\sum_{k=1}^{K} R_k = R.
\end{equation}
Each sub-message \( W_k \) (\( k \in\{ 1, 2, \ldots, K \} \)) is further split into two independent parts \( W_{k,R} \in \mathcal{W}_{k,R} \) and \( W_{k,I} \in \mathcal{W}_{k,I} \), satisfying
\begin{equation}\label{w3}
|\mathcal{W}_{k,R}| = 2^{NR_{k,R}}, \quad |\mathcal{W}_{k,I}| = 2^{NR_{k,I}},
\end{equation}
\begin{equation}\label{w4}
|\mathcal{W}_{k,R}| \cdot |\mathcal{W}_{k,I}| = |\mathcal{W}_k|,
\end{equation}
and
\begin{equation}\label{w5}
R_{k,R} + R_{k,I} = R_k.
\end{equation}
From \eqref{w2} and \eqref{w5}, we conclude that
\begin{equation}\label{w6}
\sum_{k=1}^{K} R_{k,R} + \sum_{k=1}^{K} R_{k,I} = R.
\end{equation}
A unit square in the complex plane (with side length 1) is divided into \( 2^{NR_k} \) sub-regions of equal area. This is achieved by partitioning the horizontal real axis interval \([-0.5, 0.5]\) into \( 2^{NR_{k,R}} \) equal sub-intervals, and the vertical imaginary axis interval \([-0.5, 0.5]\) into \( 2^{NR_{k,I}} \) equal sub-intervals, such that
\[
{R_{k,R}} + {R_{k,I}} = {R_k}.
\]
The transmitter then maps each sub-message \( W_k \) to a complex point \( \theta_k \). Let \( W_k = W_{k,R} + jW_{k,I} \) be the complex representation of the sub-message, where \( W_{k,R} \) and \( W_{k,I} \) correspond to the real and imaginary components, respectively. The transmitter then maps each sub-message \( W_k \) to a complex point \( \theta_k \) by
{\small 	\begin{equation}\label{eq:mapping}
	\theta_k =\theta_{k,R}+j\theta_{k,I}= -\frac{1}{2} + \frac{2W_{k,R} - 1}{2 \cdot 2^{NR_{k,R}}} + j\left( -\frac{1}{2} + \frac{2W_{k,I} - 1}{2 \cdot 2^{NR_{k,I}}} \right).
	\end{equation}}%
	Since $W_k$ is uniformly drawn, it is not difficult to check that  as $N \to \infty$, $\mathbb{E}[\|\theta_k\|^{2}] \approx \frac{1}{6}$ \cite{xie}.}
{
Then, the transmitter applies the Inverse Discrete Fourier Transform (IDFT) to the complex vector $\mathbf{D}^{\scriptsize{\textcircled{\scriptsize 1}}} = [\sqrt{6P_1}\cdot\theta_1, \cdots, \sqrt{6P_{K}}\cdot\theta_K]^{T}$, i.e.,
\begin{align}
\mathbf{d}^{\scriptsize{\textcircled{\scriptsize 1}}}_{k}& = \sum_{n=1}^{K} [{\mathbf{F}}^{-1}]_{k,n} \cdot\mathbf{D}^{\scriptsize{\textcircled{\scriptsize 1}}}_{n}\\
&=\frac{1}{\sqrt{K}} \sum_{n=1}^{K}  e^{2\pi j \cdot\frac{(n-1)(k-1)}{K}} \cdot\mathbf{D}^{\scriptsize{\textcircled{\scriptsize 1}}}_{n}, \,\,\,\, k\in\{1,\dots,K\},\nonumber
\end{align}
where $\mathbf{D}^{\scriptsize{\textcircled{\scriptsize 1}}}_{n}$ is the $n$-th component of
$\mathbf{D}^{\scriptsize{\textcircled{\scriptsize 1}}}$, and
in matrix form, the above equation can be re-written as
$$\mathbf{d}^{\scriptsize{\textcircled{\scriptsize 1}}} = {\mathbf{F}}^{-1}  \mathbf{D}^{\scriptsize{\textcircled{\scriptsize 1}}},$$
where $\mathbf{F}$ is the normalized DFT matrix with component $\mathbf{F}_{k,n}=\frac{1}{\sqrt{K}} \cdot e^{-2\pi j \cdot\frac{(n-1)(k-1)}{K}}$, for $k\in\{1,\dots,K\}$ and $n\in\{1,\dots,K\}$,
\(\mathbf{ F}^{-1} \) is the normalized Inverse Discrete Fourier Transform (IDFT) matrix with component $[{\mathbf{F}^{-1}}]_{k,n}=\frac{1}{\sqrt{K}} \cdot e^{2\pi j \cdot\frac{(n-1)(k-1)}{K}}$,
and  $P_1,\cdots,P_K$ are the power coefficients which will be optimized later.}

{	
\emph{Coding procedure}:}

{\textbf{Block $1$} (at time $i\in \{1,\dots,L+K-1\}$): the transmitted codeword is denoted by
\begin{equation}\label{sk3-X1}
X_i = [[\mathbf{d}^{\scriptsize{\textcircled{\scriptsize 1}}}]_{K-L+2}^{K} \,\,;\,\,\mathbf{d}^{\scriptsize{\textcircled{\scriptsize 1}}}]_{i}.
\end{equation}
Once the receiver receives
\begin{equation}\label{Y1}
Y_i=\sum_{l=1}^{L} h_l\cdot X_{i-l+1}+\eta_{i},
\end{equation}
and then he transmits $Y_{i}$ back to the transmitter via noiseless feedback.
Define the circulant channel matrix $\mathbf{H}_{\text{circ}}$ of size $K\times K$ with first column $[h_1, \cdots, h_L, \underset{{K-L \,\,\text{zeros}}}{\underbrace{0,\dots,0}} ]^{T}$. The auxiliary signal $\dot{\mathbf{Y}}_1 = [Y_L, \dots, Y_{L+K-1}]^{T}$ satisfies	
{\small 	\begin{align}\label{dotY1}
	\dot{\mathbf{Y}}_1
	&=\begin{bmatrix}
		h_1 & 0   & \cdots & 0   & h_L   & \cdots & h_2 \\
		h_2 & h_1 & \ddots & \vdots & \vdots & & \vdots \\
		\vdots & h_2 & \ddots & 0   & 0   & \cdots & h_L \\
		h_L & \vdots & \ddots & h_1 & 0   & \cdots & 0 \\
		0   & h_L & & h_2 & h_1 & & \vdots \\
		\vdots & \ddots & \ddots & \vdots & h_2 & \ddots & 0 \\
		0   & \cdots & 0 & h_L & \cdots & \cdots & h_1
	\end{bmatrix}
	\mathbf{d}^{\scriptsize{\textcircled{\scriptsize 1}}}
	+\begin{bmatrix}
		\eta_L\\
		\vdots\\
		\eta_{L+K-1}
	\end{bmatrix}\nonumber\\
	&=\mathbf{H}_{\text{circ}} \mathbf{d}^{\scriptsize{\textcircled{\scriptsize 1}}} + \boldsymbol{\eta}_{1},
	\end{align}}%
	where $\boldsymbol{\eta}_1 = [\eta_L, \dots, \eta_{L+K-1}]^{T}$.
	Then the receiver applies the normalized DFT to the auxiliary signal $\dot{\mathbf{Y}}_1$, i.e.,
	\begin{align}\label{FY1}
\mathbf{F}\dot{\mathbf{Y}}_1 &\mathop{=}\limits^{\text{\text{(a)}}} \mathbf{F}\mathbf{H}_{\text{circ}}\mathbf{F}^{H}\mathbf{F}\mathbf{d}^{\scriptsize{\textcircled{\scriptsize 1}}}+ \mathbf{F}\boldsymbol{\eta}_{1} \nonumber\\
&\mathop{=}\limits^{\text{\text{(b)}}} \mathbf{\Lambda}\mathbf{D}^{\scriptsize{\textcircled{\scriptsize 1}}} + \mathbf{F}\boldsymbol{\eta}_{1}=\mathbf{\Lambda}\mathbf{D}^{\scriptsize{\textcircled{\scriptsize 1}}} + \boldsymbol{Z}_{1},
\end{align}
where (a) and (b) follow from $\mathbf{F}$ is a unitary matrix satisfying $ \mathbf{F}^{-1}=\mathbf{F}^H,\,\,\mathbf{F} \mathbf{F}^H = \mathbf{F}^H \mathbf{F} = \mathbf{I}$,
$\mathbf{\Lambda}= \text{diag}(H_1, \dots, H_K)$,
\[
{H}_{k} = \sum_{n=1}^{K}  e^{-2\pi j \cdot\frac{(n-1) (k-1)}{K}} \cdot\mathbf{{\tilde{h}}}_{n},\,\,\,\,\,\,  1 \leq k\leq K,
\]
$\mathbf{{\tilde{h}}}_{n}$ is the $n$-th component of $[h_1, \cdots, h_L, \underset{{K-L \,\,\text{zeros}}}{\underbrace{0,\dots,0}} ]^{T}$, $\mathbf{D}^{\scriptsize{\textcircled{\scriptsize 1}}}=\mathbf{F}\mathbf{d}^{\scriptsize{\textcircled{\scriptsize 1}}}$,
$\boldsymbol{Z}_{1}=\mathbf{F}\boldsymbol{\eta}_{1}$, $\boldsymbol{Z}_{1}\sim \mathcal{CN}(\mathbf{0}, \boldsymbol{\Sigma}_{Z_1})$ with covariance
\begin{equation}\label{var-z1}
\boldsymbol{\Sigma}_{Z_1} =\mathbb{E}[\mathbf{F}\boldsymbol{\eta}_{1} (\mathbf{F}\boldsymbol{\eta}_{1})^{H}]= \text{diag}(\sigma^2, \dots, \sigma^2),
\end{equation}
and the fact that \emph{any circulant matrix is diagonalized by the DFT matrix $\mathbf{F}$}.
The receiver computes his first estimation
\begin{align}\label{est1}
\hat{\boldsymbol{\theta}}_1 &= \mathbf{P}^{-1}\mathbf{\Lambda}^{-1}\mathbf{F}\dot{\mathbf{Y}}_1=
\mathbf{P}^{-1}\mathbf{\Lambda}^{-1}(\mathbf{\Lambda}\mathbf{D}^{\scriptsize{\textcircled{\scriptsize 1}}} + \boldsymbol{Z}_{1})\nonumber\\
&\mathop{=}\limits^{\text{\text{(1)}}} \boldsymbol{\theta} + \mathbf{P}^{-1}\mathbf{\Lambda}^{-1}\mathbf{Z}_1=
\boldsymbol{\theta} + \boldsymbol{\epsilon}_1,
\end{align}
where $\mathbf{P} = \text{diag}(\sqrt{6P_1}, \dots, \sqrt{6P_K})$, (1) follows from $\mathbf{D}^{\scriptsize{\textcircled{\scriptsize 1}}} = [\sqrt{6P_1}\cdot\theta_1, \cdots, \sqrt{6P_{K}}\cdot\theta_K]^{T}$,
$\boldsymbol{\theta} = [\theta_1, \dots, \theta_K]^{T}$, and
$\boldsymbol{\epsilon}_1=\mathbf{P}^{-1}\mathbf{\Lambda}^{-1}\mathbf{Z}_{1}=[\epsilon_{1,1}, \dots, \epsilon_{K,1}]^{T}$.
Here note that $\boldsymbol{\epsilon}_1 \sim \mathcal{CN}(\mathbf{0}, \boldsymbol{\Sigma}_1)$ with covariance
\begin{align}\label{var-1}
\boldsymbol{\Sigma}_1 &=\mathbb{E}[\mathbf{P}^{-1}\mathbf{\Lambda}^{-1}\mathbf{Z}_{1} (\mathbf{P}^{-1}\mathbf{\Lambda}^{-1}\mathbf{Z}_{1})^{H}]\nonumber\\
&= \text{diag}(\frac{\sigma^2}{6P_1\|H_1\|^2}, \dots, \frac{\sigma^2}{6P_K\|H_K\|^2}).
\end{align}	}
{
\textbf{Block $n$ ($ n\in \{2, \cdots, \lfloor\frac{N}{L+K-1}\rfloor \}$)}: The transmitter applies IDFT to the complex vector $\mathbf{D}^{\scriptsize{\textcircled{\scriptsize n}}} = \left[\sqrt{\tfrac{P_1}{\alpha_{1,n-1}}}\epsilon_{1,n-1}, \dots, \sqrt{\tfrac{P_K}{\alpha_{K,n-1}}}\epsilon_{K,n-1}\right]$, i.e.,
\[
\mathbf{d}^{\scriptsize{\textcircled{\scriptsize n}}} = \mathbf{F}^{-1} \mathbf{D}^{\scriptsize{\textcircled{\scriptsize n}}}.
\]
where $\alpha_{k,n-1} \triangleq \mathbb{E}[\|\epsilon_{k,n-1}\|^{2}]$.	
At time  $i \in \{(L+K-1)\cdot (n-1)+1,\cdots(L+K-1)\cdot n \}$, the transmitted signal is denoted by
\begin{equation}\label{Xi}
X_i = [(\mathbf{d}^{\scriptsize{\textcircled{\scriptsize n}}})_{K-L+2}^{K} \,\,;\,\,\mathbf{d}^{\scriptsize{\textcircled{\scriptsize n}}}]_{i-(L+K-1)\cdot (n-1)}.
\end{equation}
The receiver receives
\begin{equation}\label{Yi}
Y_i=\sum_{l=1}^{L} h_l\cdot X_{i-l+1}+\eta_{i},	
\end{equation}
and transmits $Y_{i}$ back to the transmitter via noiseless feedback.	
The receiver computes an auxiliary signal $\dot{\mathbf{Y}}_n = [Y_{(L+K-1)\cdot (n-1)+L}, \cdots, Y_{n(L+K-1)}]^{T}$, which can be expressed in matrix form as
\begin{equation}\label{dotYi}
\dot{\mathbf{Y}}_n = \mathbf{H}_{\text{circ}} \mathbf{d}^{\scriptsize{\textcircled{\scriptsize n}}}+ \boldsymbol{\eta}_n,
\end{equation}
where $\boldsymbol{\eta}_n = [\eta_{(L+K-1)(n-1)+L}, \dots, \eta_{n(L+K-1)}]^{T}$.
Then the receiver applies the DFT to the auxiliary signal $\dot{\mathbf{Y}}_n$, i.e.,	
\begin{align}\label{FYi}
\mathbf{F}\dot{\mathbf{Y}}_n &= \mathbf{F}\mathbf{H}_{\text{circ}}\mathbf{F}^{H}\mathbf{F}\mathbf{d}^{\scriptsize{\textcircled{\scriptsize n}}} + \mathbf{F}\boldsymbol{\eta}_n
= \mathbf{\Lambda} \mathbf{D}^{\scriptsize{\textcircled{\scriptsize n}}} + \mathbf{Z}_n\nonumber\\
&= \begin{bmatrix}
	H_1\sqrt{\tfrac{P_1}{\alpha_{1,n-1}}}\epsilon_{1,n-1} \\
	\vdots \\
	H_K\sqrt{\tfrac{P_K}{\alpha_{K,n-1}}}\epsilon_{K,n-1}
\end{bmatrix} + \mathbf{Z}_n,
\end{align}
where $\mathbf{Z}_n=\mathbf{F}\boldsymbol{\eta}_n$ and $\boldsymbol{Z}_{n}\sim \mathcal{CN}(\mathbf{0}, \boldsymbol{\Sigma}_{Z_n})$ with covariance
\begin{equation}\label{var-zn}
\boldsymbol{\Sigma}_{Z_n} =\mathbb{E}[\mathbf{F}\boldsymbol{\eta}_{n} (\mathbf{F}\boldsymbol{\eta}_{n})^{H}]= \text{diag}(\sigma^2, \dots, \sigma^2).
\end{equation}
The receiver updates his estimation
\begin{align}\label{update-n}
\hat{\boldsymbol{\theta}}_n &= \hat{\boldsymbol{\theta}}_{n-1} - \boldsymbol{\beta}_{n-1} \mathbf{\Lambda}^{-1}\mathbf{F}\dot{\mathbf{Y}}_n\\
&=\boldsymbol{\theta}+\boldsymbol{\epsilon}_{n-1} - \boldsymbol{\beta}_{n-1} \mathbf{\Lambda}^{-1}\mathbf{F}\dot{\mathbf{Y}}_n=\boldsymbol{\theta}+\boldsymbol{\epsilon}_{n},
\end{align}
where $\boldsymbol{\epsilon}_j=\hat{\boldsymbol{\theta}}_j-\boldsymbol{\theta}$ ($j\in \{1, \cdots, \lfloor\frac{N}{L+K-1}\rfloor \} $), $\boldsymbol{\beta}_{n-1} = \text{diag}(\beta_{1,n-1}, \dots, \beta_{K,n-1})$ with  minimum mean square estimation (MMSE) coefficients in complex field given by
\begin{equation}\label{beta-n}
\beta_{k,n-1} = \frac{\mathbb{E}[\epsilon_{k,n-1}\cdot[\overline{\mathbf{\Lambda}^{-1}\mathbf{F}\dot{\mathbf{Y}}_n]_{k}}]}{\mathbb{E}[|[\mathbf{\Lambda}^{-1}\mathbf{F}\dot{\mathbf{Y}}_n]_{k}|^{2}]}=\frac{\sqrt{P_k \alpha_{k,n-1}}}{P_k + \frac{\sigma^2}{\|H_k\|^2}},
\end{equation}
where $k\in\{1,2,...,K\}$.}

{\emph{Decoding}:
After $\Phi\triangleq \lfloor\frac{N}{L+K-1}\rfloor$ blocks of iteration, the receiver's
final estimation of $\theta_k$ is denoted by $\hat{\theta}_{k,\Phi}=\theta_{k,R}+\text{Re}(\epsilon_{k,\Phi})+j(\theta_{k,I}+\text{Im}(\epsilon_{k,\Phi}))$. Choosing the
closest midpoint to  $\theta_{k,R}+\text{Re}(\epsilon_{k,\Phi})$ and $\theta_{k,I}+\text{Im}(\epsilon_{k,\Phi})$, the receiver obtains the decoded message $\hat{W}_k$ by the mapping relationship. A decoding error for message \( W \) occurs if any  sub-message $W_k$ is decoded incorrectly.}	

{	
\emph{Error probability analysis}:
Note that
{\small 		\begin{align}\label{sk-pe}
	P_e
	&=\Pr\lbrace \bigcup_{k=1}^{{K}}(|\text{Re}(\epsilon_{k,\Phi})|\geq\frac{1}{2^{NR_{k,R}+1}}\nonumber\\
	&\qquad\qquad\qquad\qquad\qquad\bigcup |\text{Im}(\epsilon_{k,\Phi})|\geq\frac{1}{2^{NR_{k,I}+1}})\rbrace\nonumber\\
	&\leq \sum_{k=1}^{K}\Pr\lbrace|\text{Re}(\epsilon_{k,\Phi})|\geq\frac{1}{2^{NR_{k,R}+1}}\rbrace\nonumber\\
	&\qquad\qquad\qquad\qquad+\sum_{k=1}^{K}\Pr\lbrace  |\text{Im}(\epsilon_{k,\Phi})|\geq\frac{1}{2^{NR_{k,I}+1}}\rbrace\\
	&=\sum_{k=1}^{K}2\mathop{Q}\left( \frac{1}{2^{NR_{k,R}+1}\cdot\sqrt{\mathbb{E}[\text{Re}(\epsilon_{k,\Phi})^{2}]}}\right)\nonumber\\
	&\qquad\qquad\qquad+\sum_{k=1}^{K}2\mathop{Q}\left( \frac{1}{2^{NR_{k,I}+1}\cdot\sqrt{\mathbb{E}[\text{Im}(\epsilon_{k,\Phi})^{2}]}}\right).\nonumber
	\end{align}}%
	To bound the error probability $P_e$ in \eqref{sk-pe}, the variances $\mathbb{E}[\text{Re}(\epsilon_{k,\Phi})^{2}]$ and $\mathbb{E}[\text{Im}(\epsilon_{k,\Phi})^{2}]$ should be determined, see the following Lemma \ref{le-1}.	
	\begin{lemma}\label{le-1}
For $n\in\{1,2,...,\Phi\}$ and $k\in\{1,2,...,K\}$, we have
\begin{equation}\label{sigma_px}
	\mathbb{E}[\|\epsilon_{k,n}\|^2] = \frac{\sigma^2}{6P_k \|H_k\|^2} \cdot \frac{1}{(1 + \|H_k\|^2\cdot \frac{P_k}{\sigma^2})^{n-1}}
\end{equation}
and
\begin{equation}\label{sigma_p1}
	\mathbb{E}[\text{Re}(\epsilon_{k,n})^{2}] = \mathbb{E}[\text{Im}(\epsilon_{k,n})^{2}] = \frac{1}{2}\mathbb{E}[\|\epsilon_{k,n}\|^2].
\end{equation}
\end{lemma}	}	
{
\begin{IEEEproof}
Lemma \ref{le-1} can be directly proved by mathematical induction, and we omit the details here.
\end{IEEEproof}}
{
Now substituting $\mathbb{E}[\text{Re}(\epsilon_{k,\Phi})^{2}]$ and $\mathbb{E}[\text{Im}(\epsilon_{k,\Phi})^{2}]$ in Lemma \ref{le-1} to \eqref{sk-pe}, we have
{\small 		\begin{equation}\label{sk3-pe-1}
	\begin{aligned}
		P_e&\leq\sum_{k=1}^{K} 2Q\left( \frac{\sqrt{12\cdot \|H_k\|^2\cdot \frac{P_k}{\sigma^{2}}(1+\|H_k\|^2\cdot\frac{P_k}{\sigma^{2}})^{\Phi-1}}}{2^{NR_{k,R}+1}} \right)
		\\&+\sum_{k=1}^{K} 2Q\left( \frac{\sqrt{12\cdot \|H_k\|^2\cdot \frac{P_k}{\sigma^{2}}(1+\|H_k\|^2\cdot\frac{P_k}{\sigma^{2}})^{\Phi-1}}}{2^{NR_{k,I}+1}} \right).\\
	\end{aligned}
	\end{equation}}
	To ensure $P_e \leq \varepsilon$ for each $k\in \{1,\dots,K\}$, let
	{\small \begin{align}\label{sk0-pe-2}
	&2Q( \frac{\sqrt{12\cdot \|H_k\|^2\cdot \frac{P_k}{\sigma^{2}}(1+\|H_k\|^2\cdot\frac{P_k}{\sigma^{2}})^{\Phi-1}}}{2^{NR_{k,R}+1}})\\
	&=2Q( \frac{\sqrt{12\cdot \|H_k\|^2\cdot \frac{P_k}{\sigma^{2}}(1+\|H_k\|^2\cdot\frac{P_k}{\sigma^{2}})^{\Phi-1}}}{2^{NR_{k,I}+1}})=\frac{\varepsilon}{2K}.\nonumber
	\end{align}}
	From \eqref{sk0-pe-2}, the rates $R_{k,R}$ and $R_{k,I}$ can be computed as
	\begin{equation}
\begin{aligned}
	R_{k,R} &= \frac{\Phi-1}{2N}\log\left(1+\|H_k\|^2\cdot\frac{P_k}{\sigma^{2}}\right)\nonumber \\
	&\qquad\qquad\qquad- \frac{1}{2N}\log\left(\frac{\Xi}{12 \cdot \|H_k\|^{2}\cdot\frac{P_k}{\sigma^{2}}}\right), \\
	R_{k,I} &= \frac{\Phi-1}{2N}\log\left(1+\|H_k\|^2\cdot\frac{P_k}{\sigma^{2}}\right)\\
	&\qquad\qquad\qquad - \frac{1}{2N}\log\left(\frac{\Xi}{12 \cdot \|H_k\|^{2}\cdot\frac{P_k}{\sigma^{2}}}\right),
\end{aligned}
\end{equation}
where $\Xi = 4\left[Q^{-1}\left(\frac{\varepsilon}{4K}\right)\right]^2$ and $\Phi = \lfloor\frac{N}{L+K-1}\rfloor$.	
Then, according to \eqref{w6}, the achievable rate $\mathcal{R}_\text{L-Path}(N,\varepsilon)$ of the model in Figure~\ref{path-1} is given by
\begin{equation}\label{sk3-Rxfd}
\begin{aligned}
	\mathcal{R}_\text{L-Path}(N,\varepsilon) &= \sum_{k=1}^{K} R_{k,R} + \sum_{k=1}^{K} R_{k,I} \\
	&= \sum_{k=1}^{K} \left( \frac{\Phi-1}{N}\log\left(1+\|H_k\|^2\cdot\frac{P_k}{\sigma^{2}}\right) \right.\\	
	&\qquad\qquad 	-\left. \frac{1}{N}\log\left(\frac{\Xi}{12 \cdot \|H_k\|^{2}\cdot\frac{P_k}{\sigma^{2}}}\right) \right),
\end{aligned}
\end{equation}
where $\Xi = 4\left[Q^{-1}\left(\frac{\varepsilon}{4K}\right)\right]^2$ and $\Phi = \lfloor\frac{N}{L+K-1}\rfloor$. Then optimizing
$P_1,\cdots,P_K$ by the water-filling algorithm, namely,			
\begin{equation}\label{P}
P_k=\max(q-\frac{\sigma^2}{\|H_k\|^{2}},0),\,\,\,\, 1\leq k\leq K,
\end{equation}
where $q$ is chosen to satisfy
\begin{equation}
\sum_{k=1}^{K}\max(q-\frac{\sigma^2}{\|H_k\|^{2}},0)=KP,
\end{equation}
and maximizing $\mathcal{R}_\text{L-Path}(N,\varepsilon)$ over $K\in \{L,\cdots,N-L+1\}$, the proof is completed.}

\section{Conclusion and Further Work}\label{Conclusion}

In the literature, the classical SK scheme has already been shown to be an excellent finite blocklength coding scheme. However, this scheme is extremely sensitive to the CSI and channel output feedback, which indicates that extending it to the practical wireless fading channels is still a big challenge.
In this paper, we extend the classical SK scheme to the quasi-static fading channel with I-CSIT and QFC, and the two-ray channel with I-CSIT and QFC. {Besides this, for arbitrary multi-path fading channel with feedback, we also present an SK-type scheme for such a model, which 
transforms the time domain channel into a frequency domain MIMO channel, and adopts the SK-type scheme for the fading MIMO channels. }

{One possible future work is to extend our scheme to the fading MIMO channel with I-CSIT and QFC.
The challenge to this extension is that if the transmitter only knows an estimation of the CSI matrix, traditional singular value decomposition (SVD) on the CSI matrix may not work since we do not know the property of the transmitter's estimation about the CSI martrix.
}

{Another possible future work is to extend our scheme to the fading channel with noisy feedback channel and quantized feedback signal. In such a case, since the feedback channel noise is not known by the receiver, how to eliminate the impact of this noise on the performance of the SK-type scheme is challenging. }

{Besides this, investigating how to design SK-type scheme with the consideration of hardware impairments \cite{re1,re2,re3} in practical wireless systems is also interesting and deserves to be our future work.}

\renewcommand{\theequation}{A\arabic{equation}}
\appendices

\section{Proof of Lemma \ref{sk2-lemma1} in Theorem \ref{th2}}\label{b-0}

Similar to that of the proof of Theorem \ref{th1}, first, we show that the decoding error probability $\Pe$ of the original system is no larger than that of the coupled system, namely,
{\small \begin{equation}\label{sk2-app-1}
P_e\leq \sum_{i=2}^{N-1}\Pr\{E_i'\}+\Pr\{E_N'\}.
\end{equation}}%
Since the proof is similar to that of Theorem \ref{th1}, we omit the detailed proof here.

In order to further bound (\ref{sk2-app-1}), it remains to determine $\Pr\{E_i'\}$ and $\Pr\{E_N'\}$, see the detail below.

\emph{ 1) An upper bound on $\Pr\{E_i'\}$}:	For $i\in\{2,\cdots,N-1\}$, similar to (\ref{sk1-gamma1-1}), we have
{\small \begin{equation}\label{sk2-gamma-1}	
\begin{aligned}
	\Pr\lbrace E_i^{'}\rbrace
	&\leq \Pr\{\arrowvert\gamma_i\cdot{\epsilon}_i^{'}\arrowvert+\sigma_z\geq\sqrt{3\widetilde{P}}\}.	
\end{aligned}									
\end{equation}}%
To further bound $\Pr\{E_i'\}$ of (\ref{sk2-gamma-1}), we need to determine
${\epsilon}_i^{'}$ and
{\small \begin{equation}\label{sk2-alpha_i}	
\alpha_i' = \mathbb{E}[\epsilon_i^{'2}]
\end{equation}}%
of the coupled system first, where
{\small 	\begin{equation}\label{sk2-ep1}	
\epsilon_{i+1}' = \epsilon_{i}' - \beta_{i} \cdot \dot{Y}_{i+1}'.
\end{equation}}%
To determine ${\epsilon}_i^{'}$ and $\alpha_i'$, it remains to
calculate the terms $\dot{Y}_{i+1}'$ and $\beta_{i}$ of (\ref{sk2-ep1}), see the detail below.

\emph{Calculation of {\small $\dot{Y}_{i+1}'$}}:
First, recall that in the original system, based on  (\ref{sk2-y3}) and (\ref{sk2-yi}), we have
{\small 	 	\begin{align}
&\begin{aligned}
	{Y}_{3}=\alpha h_1\cdot\sgn{h_1h_2}\cdot\mathbb{M}_{\widetilde{d}}[\gamma_{2}\cdot\epsilon_{2}+Z_2]+\eta_{3},
\end{aligned}\\
&\begin{aligned}
	{Y}_{i}=&\alpha h_1\cdot[\sgn{h_1h_2}]^{i-2}\cdot\mathbb{M}_{\widetilde{d}}[\gamma_{i-1}\cdot \epsilon_{i-1}+Z_{i-1}]
	\\&+\alpha h_2\cdot[\sgn{h_1h_2}]^{i-3}\cdot\mathbb{M}_{\widetilde{d}}[\gamma_{i-2}\cdot \epsilon_{i-2}+Z_{i-2}]+\eta_i,
\end{aligned}\nonumber
\end{align}}%
for {\small $i \in \left\lbrace 4,\ldots,N\right\rbrace$}, and the auxiliary signal $\dot{Y}_i$ of (\ref{sk2-aux1}) and (\ref{sk2-aux2}) can be re-written by
{\small 	\begin{align}
&\begin{aligned}
	\dot{Y}_{3}=&\alpha h_1\cdot\sgn{h_1h_2}(\mathbb{M}_{\widetilde{d}}[\gamma_{2}\cdot\epsilon_{2}+Z_2]-Z_2)+\eta_{3}\\
\end{aligned}\\
&\begin{aligned}
	\dot{Y}_i=&\alpha h_1\cdot[\sgn{h_1h_2}]^{i-2}(\mathbb{M}_{\widetilde{d}}[\gamma_{i-1}\cdot \epsilon_{i-1}+Z_{i-1}]- Z_{i-1})\\
	&+\alpha h_2\cdot[\sgn{h_1h_2}]^{i-3}\mathbb{M}_{\widetilde{d}}[\gamma_{i-2}\cdot \epsilon_{i-2}+Z_{i-2}]\nonumber\\
	&-\alpha h_{2}[\sgn{h_1h_2}]^{i-3}\cdot (Z_{i-2}+\beta_{i-2}\cdot\dot{Y}_{i-1})+\eta_i.
\end{aligned}\nonumber	
\end{align}}%
Next, note that all modulo-lattice functions in the original system are removed in the coupled system, hence we have
{\small \begin{align}\label{sk2-dotY}
&\begin{aligned}
	\dot{Y}_{3}^{'}= \alpha \cdot\sgn{h_1h_2}\cdot h_{1}\cdot\gamma_{2}\cdot{\epsilon}_{2}^{'}+\eta_{3},
\end{aligned}\\
&\begin{aligned}
	&\dot{Y}_i^{'}
	\mathop{=}\limits^{\text{(a)}} \alpha\cdot [\sgn{h_1h_2}]^{i-2}\cdot h_{1}\cdot\gamma_{i-1}\cdot{\epsilon}_{i-1}'\\
	&\qquad+[\sgn{h_1h_2}]^{i-3}\cdot h_{2}\cdot\gamma_{i-2}\cdot{\epsilon}_{i-1}'
	+\eta_{i},
\end{aligned}\nonumber	
\end{align}}%
where $i \in \left\lbrace 4,\ldots,N\right\rbrace$, and $\text{(a)}$ follows from (\ref{sk2-ep1}).

\emph{Calculation of $\beta_{i}$}: Note that $\beta_{i}$ is MMSE coefficient for both original and coupled systems, and it is given by
{\small 	\begin{equation}\label{sk2-beta}
\beta_{i} = \frac{\mathbb{E}[\epsilon_{i}' \cdot \dot{Y}_{i+1}']}{\mathbb{E}[(\dot{Y}_{i+1}')^2]}.
\end{equation}}%
Since $\beta_{i}$ is correlated with $\epsilon_{i}'$, currently it cannot be calculated, and we will determine it later.

Based on the determination of $\dot{Y}_{i+1}'$ and $\beta_{i}$, $\alpha_i'$ of (\ref{sk2-alpha_i}) can be further re-written as
{\footnotesize 	\begin{align}
&\alpha_{2}'=\frac{\sigma^2}{ 12P\cdot(h_1^{2}+h_2^{2})},\label{sk2-alpha2-2}\\
&\alpha_{3}'=\alpha_{2}'-\frac{(\mathbb{E}[\epsilon_{2}^{'}\cdot\dot{Y}_{3}^{'}])^{2}}{\mathbb{E}[\dot{Y}_{3}^{'2}]}
=({1+\frac{h_1^2\cdot\alpha^2\cdot\gamma_2^2\cdot\alpha_{2}'}{\sigma^2}})^{-1}\cdot\alpha_{2}', \label{sk2-alpha2-3}\\
&\alpha_{i}'=\alpha_{i-1}'-\frac{(\mathbb{E}[\epsilon_{i-1}^{'}\dot{Y}_{i}^{'}])^{2}}{\mathbb{E}[\dot{Y}_{i}^{'2}]}
({1+\frac{\alpha^2\Xi_{i-1}^2\alpha_{i-1}'}{\sigma^2}})^{-1}\cdot\alpha_{i-1}',\label{sk2-alpha2-i}
\end{align}}%
where {\small $i\in \{4,\cdots,N\}$} and {\small $\Xi_{i-1}=[\sgn{h_1h_2}]^{i-2}\cdot h_{1}\cdot\gamma_{i-1}+[\sgn{h_1h_2}]^{i-3}\cdot h_{2}\cdot\gamma_{i-2}$}.

From (\ref{sk2-dotY}), we conclude that $\dot{Y}_i'$ is Gaussian distributed, leading to $\epsilon_{i}^{'}$ of (\ref{sk2-ep1}) also be Gaussian distributed. Hence
(\ref{sk2-gamma-1}) can be further re-written as
{\footnotesize \begin{equation}\label{sk2-gamma-2}	
\Pr\lbrace E_i^{'}\rbrace\!\leq\! \Pr\{\arrowvert\gamma_i{\epsilon}_i^{'}\arrowvert\!+\!\sigma_z \! \geq\!\sqrt{3\widetilde{P}}\}
\!=\!2Q(\!\frac{\sqrt{3\widetilde{P}}\!- \!\sigma_z }{ \sqrt{\gamma_i^2\alpha_i'}} \!).								
\end{equation}}%
Analogous to the proof of Theorem \ref{th1},  instead of $\alpha_i'$, $ \alpha_{i}'(H_1,H_2)$ depending on $\hat{h_1}$, $\hat{h_2}$ and $D$ should be adopted by the transmitter, namely,
{\footnotesize 	\begin{align}
&\alpha_{2}'(H_{1},H_{2})=\frac{\sigma^2}{ 12P\cdot(H_1^2+H_2^2)},\label{sk2-alpha3-2}\\
&\alpha_{3}'(H_{1},H_{2})=\alpha_{2}'(H_{1},H_{2})
({1+\frac{H_1^2\alpha^2\gamma_2^2\alpha_{2}'(H_{1},H_{2})}{\sigma^2}})^{-1}, \label{sk2-alpha3-3}\\
&\alpha_{i}'(H_{1},H_{2})=\alpha_{i-1}'(H_{1},H_{2})\nonumber\\
&\,\, \times({1+\frac{\alpha^2\cdot(|H_{1}|\gamma_{i-1}+|H_{2}|\gamma_{i-2})^2\cdot \alpha_{i-1}'(H_{1},H_{2})}{\sigma^2}})^{-1},\label{sk2-alpha3-i}
\end{align}}%
where {\small $i\in \{4,\cdots,N\}$}, {\small $H_1=\max(\mid \hat{h}_1\mid-{D},0)$} and {\small $H_2=\max(\mid \hat{h}_2\mid-{D},0)$}.

Next, since
{\small \begin{equation}\label{sk2-alpha-vs}
\alpha_{j}'(H_1,H_2) \geq \alpha_{j}'
\end{equation}}%
for any $j \in \{2,\ldots,N\} $, and $Q(x)$-function monotonically decreases as $x$ increases, we have	
{\footnotesize \begin{equation}\label{sk2-gamma-3}	
\Pr\lbrace E_i^{'}\rbrace\leq 2Q(\frac{\sqrt{3\widetilde{P}}-\sigma_z }{ \sqrt{\gamma_i^2\cdot\alpha_i'}} ) \leq 2Q(\frac{\sqrt{3\widetilde{P}} - \sigma_z}{\sqrt{\gamma_i^2 \cdot \alpha_{i}'(H_{1},H_{2})}} ).									
\end{equation}}%

\emph{2) An upper bound on $\Pr\{E_N'\}$}: Note that
{\small \begin{align}\label{sk2-Pe1}
&\Pr\{E_{N}'\}=\Pr\lbrace \epsilon_{N}' \notin [ -\frac{1}{2^{NR+1}},\frac{1}{2^{NR+1}}) \rbrace\\
&=2\mathop{Q}( \frac{1}{2^{NR+1}\sqrt{\alpha_{N}'}}) 	
\leq 2\mathop{Q}( \frac{1}{2^{NR+1}\sqrt{\alpha_{N}'(H_1,H_2)}}).\nonumber
\end{align}}%
Substituting (\ref{sk2-gamma-3}) and (\ref{sk2-Pe1}) into (\ref{sk2-app-1}), we get
{\footnotesize 	\begin{align}\label{sk2-pe3}	
&P_e\leq\sum_{i=2}^{N-1}\Pr\{E_i'\}+\Pr\{E_{N}'\}\\
&\leq\sum_{i=2}^{N-1}2Q(\frac{\sqrt{3\widetilde{P}}-\sigma_z}{\sqrt{\gamma_i^2 \alpha_{i}'(H_1,H_2)}} )
\!+\!2\mathop{Q}( \frac{1}{2^{NR+1}\sqrt{\alpha_{N}'(H_1,H_2)}}).\nonumber				
\end{align}	}%
The proof of Lemma \ref{sk2-lemma1} in Theorem \ref{th2} is completed.						

\end{document}